\pdfoutput=1

\documentclass[10pt,a4paper]{article}

\usepackage{graphicx}

\usepackage{url,algorithm}
\usepackage[titletoc]{appendix}
\usepackage{amsmath}
\usepackage{amssymb}
\usepackage{subfigure}
\usepackage{euscript}
\usepackage{verbatim}
\usepackage{cite}
\usepackage{color}
\usepackage{multirow}
\usepackage{nameref}
\usepackage{subfigure}
\usepackage{pgf}
\usepackage{tikz}
\usetikzlibrary{calc}
\usetikzlibrary{arrows,automata}
\usepackage[latin1]{inputenc}
\usepackage{pgfplots}
\usepackage{diagbox}
\usepackage{calligra}
\usetikzlibrary{calc,positioning}
\usetikzlibrary{shapes}

\newtheorem{remark}{Remark}{\normalfont}{\normalfont}

\usepackage{enumitem}
\renewcommand{\theenumi}{\arabic{enumi}}

\renewcommand{\theenumii}{\arabic{enumii}}

\renewcommand{\theenumiii}{\arabic{enumiii}}

\newcommand{\smallmat}[1]{\left[ \begin{smallmatrix}#1 \end{smallmatrix} \right]}
\DeclareMathOperator*{\argmin}{argmin}

\title{\Large Cloud-aided collaborative estimation by ADMM-RLS algorithms for connected vehicle prognostics \\	
\large{\vspace{0.5cm}Technical Report TR-2017-01 \footnotetext{\hspace{-0.6cm}
		The results in the report have been partially presented in a paper submitted to ACC 2018.\\
}} \ \\
}
\author{Valentina Breschi\thanks{Valentina Breschi and Alberto Bemporad are with the IMT School for Advanced Studies Lucca, Piazza San Francesco 19, 55100 Lucca, Italy.
		{\tt\small valentina.breschi@imtlucca.it}; {\tt\small alberto.bemporad@imtlucca.it}} , Ilya Kolmanovsky\thanks{Ilya Kolmanovsky is with the Department of Aerospace Engineering, University of Michigan, Ann Arbor, MI 48109, USA. {\tt\small ilya@umich.edu}} , Alberto Bemporad$^*$%
}
\begin{document}

\maketitle

\tikzstyle{communication block} = [draw, fill=gray!15, rectangle, rounded corners,
minimum height=6em, minimum width=6em]
\tikzstyle{collect block} = [draw, fill=red!20, rectangle, rounded corners,
minimum height=3em, minimum width=6em]
\tikzstyle{operational block} = [draw, fill=blue!20, rectangle, rounded corners,
minimum height=3em, minimum width=6em]
\tikzstyle{cloud node} = [fill=gray!5, rectangle, 
minimum height=3em, minimum width=6em]
\tikzstyle{broadcasting block} = [draw, fill=white!5!green!10, rectangle, rounded corners,
minimum height=3em, minimum width=6em]
\tikzstyle{car block main} = [fill=blue!4, rectangle, rounded corners,
minimum height=3em, minimum width=6em]
\tikzstyle{car block} = [fill=white, rectangle, 
minimum height=3em, minimum width=6em]
\tikzstyle{dot block} = [fill=white, rectangle, 
minimum height=3em, minimum width=6em]
\tikzstyle{aid node} = [coordinate]

\begin{abstract}
	As the connectivity of consumer devices is rapidly growing and cloud computing technologies are becoming more widespread, cloud-aided techniques for parameter estimation can be designed to exploit the theoretically unlimited storage memory and computational power of the \textquotedblleft cloud\textquotedblright, while relying on information provided by multiple sources.\\
	With the ultimate goal of developing monitoring and diagnostic strategies, this report focuses on the design of a Recursive Least-Squares (RLS) based estimator for identification over a group of devices connected to the \textquotedblleft cloud\textquotedblright. The proposed approach, that relies on Node-to-Cloud-to-Node (N2C2N) transmissions, is designed so that: ($i$) estimates of the unknown parameters are computed locally and ($ii$) the local estimates are refined on the cloud. The proposed approach requires minimal changes to local (pre-existing) RLS estimators.  
\end{abstract}

\section{Introduction}
\begin{figure}
	\hspace{0cm}
	\centering
	\resizebox{8cm}{8cm}{
		\begin{tikzpicture}[auto, node distance=2cm,>=latex']
		\node [draw, cloud, fill=gray!3, cloud puffs=10, aspect=2, inner sep=0cm] (Main) {
			\begin{tikzpicture}
			\node [communication block] (A) {\begin{tabular}{c} \underline{\Large{Communication Layer}} \vspace{0.2cm}\\
				\begin{tikzpicture}
				\node [collect block] (A1) {\Large{Collect information}};
				\node [broadcasting block, left of=A1, node distance=5cm] (D) {\begin{tabular}{c}\Large{Broadcast} \\
					\Large{Global Updates}\end{tabular}};
				\end{tikzpicture}\end{tabular}};
		\node [operational block, above of=A, node distance=3cm] (B) {\Large{Global Updates}};
		\node [aid node, left of=A, node distance=2.9cm] (Aidl1){};
		\node [aid node, above of=Aidl1, node distance=0.3cm] (Aidl2){};
		\draw [->,thick, black!70!green,bend right] (B) to[] node[name=connection2] {} (Aidl2);
		\node [aid node, right of=A, node distance=2.9cm] (Aidr1){};
		\node [aid node, above of=Aidr1, node distance=0.3cm] (Aidr2){};
		\draw [->,thick, black!50!red,bend right] (Aidr2) to[] node[name=connection1] {} (B);
		\node [cloud node, above right of=connection1, node distance=1.5cm] (C) {\textcolor{black}{\underline{\textbf{\Large{CLOUD}}}}};
\end{tikzpicture}};
\node [car block main, below of= A,node distance=9cm] (1) {
	\begin{tikzpicture}[->,>=stealth',shorten >=1pt,auto,node distance=2.8cm,
	semithick]
	\tikzstyle{every state}=[fill=red,draw=none,text=white]
	
	\node[] (Ain)                    {\includegraphics[scale=0.5]{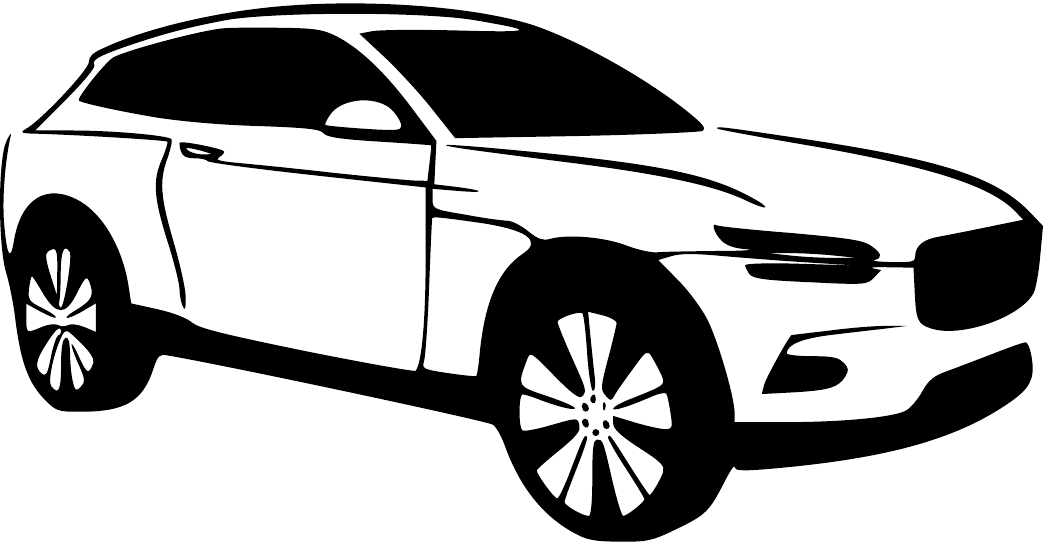}};
	\node[operational block, below of=1, node distance=1.5cm]         (Bin) [below of=Ain]       {\Large{Local Updates}};
	\node[aid node, left of=Bin, node distance=1.6cm]         (Bin1) {};
	\node[aid node, right of=Bin, node distance=1.65cm]         (Bin2) {};
	
	\path (Ain) edge[bend right,thick]         node {} (Bin1)
	(Bin2) edge [bend right,thick] node {} (Ain);
	\end{tikzpicture}
};
\node [dot block, right of=1, node distance=4cm] (1r) {$\boldsymbol{\cdots}\boldsymbol{\cdots}$};
\node [car block, right of=1r, node distance=3cm] (2) {\includegraphics[scale=0.3]{car-eps-converted-to}};
\node [dot block, left of=1, node distance=4cm] (1l) {$\boldsymbol{\cdots}\boldsymbol{\cdots}$};
\node [car block, left of=1l, node distance=3cm] (3) {\includegraphics[scale=0.3]{car-eps-converted-to}};
\node [aid node, below of=A, node distance=2.35cm] (baseaid1){};
\node [aid node, right of=baseaid1, node distance=0.1cm] (aid11){};
\node [aid node, left of=baseaid1, node distance=0.1cm] (aid12){};
\node [aid node, right of=baseaid1, node distance=1cm] (baseaid1r){};
\node [aid node, right of=baseaid1r, node distance=0.1cm] (aid1r1){};
\node [aid node, left of=baseaid1r, node distance=0.1cm] (aid1r2){};
\node [aid node, left of=baseaid1, node distance=1cm] (baseaid1l){};
\node [aid node, right of=baseaid1l, node distance=0.1cm] (aid1l1){};
\node [aid node, left of=baseaid1l, node distance=0.1cm] (aid1l2){};
\node [aid node, above of=1, node distance=2.7cm] (baseaidcar1){};
\node [aid node, right of=baseaidcar1, node distance=0.1cm] (aidcar11){};
\node [aid node, left of=baseaidcar1, node distance=0.1cm] (aidcar12){};
\node [aid node, above of=2, node distance=1cm] (baseaidcar2){};
\node [aid node, right of=baseaidcar2, node distance=-0.9cm] (aidcar21){};
\node [aid node, left of=baseaidcar2, node distance=1.2cm] (aidcar22){};
\node [aid node, above of=3, node distance=1cm] (baseaidcar3){};
\node [aid node, right of=baseaidcar3, node distance=1.2cm] (aidcar31){};
\node [aid node, left of=baseaidcar3, node distance=-0.9cm] (aidcar32){};
\draw [->,dashed,very thick,black!50!red] (aidcar11) to[] node[auto] {} (aid11);
\draw [->,dashed,very thick,black!50!red] (aidcar21) to[] node[auto] {} (aid1r1);
\draw [->,dashed,very thick,black!50!red] (aidcar31) to[] node[auto] {} (aid1l1);
\draw [->,dashed,very thick, black!70!green] (aid1l2) to[] node[auto] {} (aidcar32);
\draw [->,dashed,very thick, black!70!green] (aid1r2) to[] node[auto] {} (aidcar22);
\draw [->,dashed,very thick, black!70!green] (aid12) to[] node[auto] {} (aidcar12);
\end{tikzpicture}}
\caption{Cloud-connected vehicles.}
\label{Fig:toy_example}
\vspace{-0.5cm}
\end{figure}

With the increasing connectivity between devices, the interest in distributed solutions for estimation \cite{Olfati-Saber2007}, control \cite{Garin2010} and machine learning \cite{Forero2010} has been rapidly growing. In particular, the problem of parameter estimation over networks has been extensively studied, especially in the context of Wireless Sensor Networks (WSNs). The methods designed to solve this identification problem can be divided into three groups: incremental approaches \cite{Lopes2007}, diffusion approaches \cite{Cattivelli2008} and consensus-based distributed strategies \cite{Mateos2009}. Due to the low communication power of the nodes in WSNs, research has mainly been devoted to obtain fully distributed approaches, i.e. methods that allow exchanges of information between neighbor nodes only. Even though such a choice enables to reduce multi-hop transmissions and improve robustness to node failures, these strategies allows only neighbor nodes to communicate and thus to reach consensus. As a consequence, to attain consensus on the overall network, its topology has to be chosen to enable exchanges of information between the different groups of neighbor nodes.\\
At the same time, with recent advances in cloud computing \cite{Mell2011} it has now become possible to acquire and release resources with minimum effort so that each node can have on-demand access to shared resources, theoretically characterized by unlimited storage space and computational power. This motivates to reconsider the approach towards a more centralized strategy where some computations are performed at the node level, while the most time and memory consuming ones are executed \textquotedblleft on the cloud\textquotedblright. This requires the communication between the nodes and a fusion center, i.e. the \textquotedblleft cloud\textquotedblright, where the data gathered from the nodes are properly merged.\\
Cloud computing has been considered for automotive vehicle applications in \cite{Li2016}-\cite{Li2017} and \cite{Ozatay2014}. As motivating example for another possible automotive application, consider a vehicle fleet with vehicles connected to the \textquotedblleft cloud\textquotedblright \ (see~\figurename{~\ref{Fig:toy_example}}). 
In such a setting, measurements taken on-board of the vehicles can be used for cloud-based diagnostics and prognostics purposes. In particular, the measurements can be used to estimate parameters that may be common to all vehicles, such as parameters in components wear models or fuel consumption models, and parameters that may be specific to individual vehicles.  References \cite{Taheri2016} and \cite{howell2010brake} suggest potential applications of such approaches for prognostics of automotive fuel pumps and brake pads. Specifically, the component wear rate as a function of the workload (cumulative fuel flow or energy dissipated in the brakes) can be common to all vehicles or at least to all vehicles in the same class.\\
\noindent A related distributed diagnostic technique has been proposed in \cite{Boem2012}. However it relies on a fully-distributed scheme, introduced to reduce long distance transmissions and to avoid the presence of a \textquotedblleft critic\textquotedblright \ node in the network, i.e. a node whose failure causes the entire diagnostic strategy to fail.\\

In this report a centralized approach for recursive estimation of parameters in the least-squares sense is presented. The method has been designed under the hypothesis of ($i$) ideal transmission, i.e. the information exchanged between the cloud and the nodes is not corrupted by noise, and the assumption that ($ii$) all the nodes are described by the same model, which is supposed to be known a priori. Differently from what is done in many distributed estimation methods (e.g. see~\cite{Mateos2009}), where the nodes estimate common unknown parameters, the strategy we propose allows to account for more general consensus constraint. As a consequence, for example, the method can be applied to problems where only a subset of the unknowns is common to all the nodes, while other parameters are purely local, i.e. they are different for each node.\\
Our estimation approach is based on defining a separable optimization problem which is then solved through the Alternating Direction Method of Multipliers (ADMM), similarly to what has been done in \cite{Mateos2009} but in a somewhat different setting. As shown in \cite{Mateos2009}, the use of ADMM leads to the introduction of two time scales based on which the computations have to be performed. In particular, the local time scale is determined by the nodes' clocks, while the cloud time scale depends on the characteristics of the resources available in the center of fusion and on the selected stopping criteria, used to terminate the ADMM iterations.\\
The estimation problem is thus solved through a two-step strategy. In particular: ($i$) local estimates are recursively retrieved by each node using the measurements acquired from the sensors available locally; ($ii$) global computations are performed to refine the local estimates, which are supposed to be transmitted to the cloud by each node. Note that, based on the aforementioned characteristics, back and forth transmissions to the cloud are required. A transmission scheme referred to as Node-to-Cloud-to-Node (N2C2N) is thus employed.\\
The main features of the proposed strategies are: ($i$) the use of recursive formulas to update the local estimates of the unknown parameters; ($ii$) the possibility to account for the presence of both purely local and global parameters, that can be estimated in parallel;  ($iii$) the straightforward integration of the proposed techniques with pre-existing Recursive Least-Squares (RLS) estimators already running on board of the nodes.\\

The report is organized as follows. In Section~\ref{Sec:ADMM_base} ADMM is introduced, while in Section~\ref{Sec:prob_for} is devoted to the statement of the considered problem. The approach for collaborative estimation with full consensus is presented in Section~\ref{Sec:1}, along with the results of simulation examples that show the effectiveness of the approach and its performance in different scenarios. In Section~\ref{Sec:2} and Section~\ref{Sec:3} the methods for collaborative estimation with partial consensus and for constrained collaborative estimation with partial consensus are described, respectively. Results of simulation examples are also reported. Concluding remarks and directions for future research are summarized in Section~\ref{Sec:Conclusions}.

\subsection{Notation}
Let $\mathbb{R}^{\mathsf{n}}$ be the set of real vectors of dimension $\mathsf{n}$ and $\mathbb{R}^{+}$ be the set of positive real number, excluding zero. Given a set $\mathcal{A}$, let $\breve{\mathcal{A}}$ be the complement of $\mathcal{A}$. Given a vector $a\in\mathbb{R}^{\mathsf{n}}$, $\|a\|_2$ is the Euclidean norm of $a$. Given a matrix $A\in\mathbb{R}^{\mathsf{n}\times \mathsf{p}}$, $A'$ denotes the transpose of $A$. Given a set $\mathcal{A}$, let $\mathcal{P}_{\mathcal{A}}$ denote the Euclidean projection onto $\mathcal{A}$. Let $I_\mathsf{n}$ be the identity matrix of size $\mathsf{n}$ and $0_\mathsf{n}$ be an $\mathsf{n}$-dimensional column vector of ones.

\section{Alternating Direction Method of Multipliers} \label{Sec:ADMM_base}
The Alternating Direction Method of Multipliers (ADMM) \cite{ADMMBoyd} is an algorithm tailored for problems in the form
\begin{equation}\label{eq:ADMMprob}
\begin{aligned}
&\mbox{minimize } && f(\theta)+\\
&\mbox{subject to } && A\theta+Bz=c,
\end{aligned}
\end{equation}
where $\theta \in \mathbb{R}^{n_{\theta}}$, $z \in \mathbb{R}^{n_{z}}$, $f:\mathbb{R}^{n_{\theta}} \rightarrow \mathbb{R}\cup\{+\infty\}$ and $g: \mathbb{R}^n_{z} \rightarrow \mathbb{R}\cup\{+\infty\}$ are closed, proper, convex functions and $A \in \mathbb{R}^{p\times n_{\theta}}$, $B \in \mathbb{R}^{p \times n_{z}}$, $c \in \mathbb{R}^{p}$.\\

To solve Problem~\eqref{eq:ADMMprob}, the ADMM iterations to be performed are
\begin{align}
& \theta^{(k+1)}=\underset{\theta}{\argmin} \ \mathcal{L}(\theta,z^{(k)},\delta^{(k)}), \label{admmStep:1} \\
& z^{(k+1)}=\underset{z}{\argmin} \  \mathcal{L}(\theta^{(k+1)},z,\delta^{(k)}), \label{admmStep:2}\\
& \delta^{(k+1)}=\delta^{(k)}+\rho(A\theta^{(k+1)}+Bz^{(k+1)}-c), \label{admmStep:3} 
\end{align}
where $k \in \mathbb{N}$ indicates the ADMM iteration, $\mathcal{L}$ is the augmented Lagrangian associated to \eqref{eq:ADMMprob}, i.e.
\begin{equation}\label{eq:aLag1}
\mathcal{L}(\theta,z,\delta)=f(\theta)+g(z)+\delta'\left(A\theta+Bz-c\right)+\frac{\rho}{2}\left\|A\theta+Bz-c\right\|_{2}^{2},
\end{equation}
$\delta \in \mathbb{R}^{p}$ is the Lagrange multiplier and $\rho \in \mathbb{R}^{+}$ is a tunable parameter (see~\cite{ADMMBoyd} for possible tuning strategies). Iterations \eqref{admmStep:1}-\eqref{admmStep:3} have to be run until a stopping criteria is satisfied, e.g. the maximum number of iterations is attained.\\ 

It has to be remarked that the convergence of ADMM to high accuracy results might be slow (see~\cite{ADMMBoyd} and references therein). However, the results obtained with a few tens of iterations are usually accurate enough for most of applications. For further details, the reader is referred to \cite{ADMMBoyd}.

\subsection{ADMM for constrained convex optimization} 
Suppose that the problem to be addressed is
\begin{equation}\label{eq:constr_probl}
\begin{aligned}
& \min_{\theta} &&\  f(\theta)\\
& \mbox{s.t. }&& \theta \in \mathcal{C},
\end{aligned}
\end{equation}
with $\theta \in \mathbb{R}^{n_\theta}$, $f: \mathbb{R}^{n_{\theta}}\rightarrow \mathbb{R}\cup\{+\infty\}$ being a closed, proper, convex function and $\mathcal{C}$ being a convex set, representing constraints on the parameter value.\\
As explained in \cite{ADMMBoyd}, \eqref{eq:constr_probl} can be recast in the same form as \eqref{eq:ADMMprob} through the introduction of the auxiliary variable $z \in \mathbb{R}^{n_{\theta}}$ and the indicator function of set $\mathcal{C}$, i.e.
\begin{equation}\label{eq:ind_func}
g(z)=\begin{cases} 0 &\mbox{ if } z \in \mathcal{C}\\
+\infty &\mbox{ otherwise}
\end{cases}.
\end{equation} 
In particular, \eqref{eq:constr_probl} can be equivalently stated as 
\begin{equation}\label{eq:constr_prob2}
\begin{aligned}
& \min_{\theta,z} && f(\theta)+g(z)\\
& \mbox{s.t.} && \theta-z=0.
\end{aligned}
\end{equation}
Then, the ADMM scheme to solve \eqref{eq:constr_prob2} is
\begin{align} 
\theta^{(k+1)}&=\underset{\theta}{\argmin} \ \mathcal{L}(\theta,z^{(k)},\delta^{(k)}),\\
z^{(k+1)}&=\mathcal{P}_{\mathcal{C}}(\theta^{(k+1)}+\delta^{(k)}),\\
\delta^{(k+1)}&=\delta^{(k)}+\rho(\theta^{(k+1)}-z^{(k+1)})
\end{align}
with $\mathcal{L}$ equal to
\begin{equation*}
\mathcal{L}(\theta,z,\delta)=f(\theta)+g(z)+\delta'(\theta-z)+\frac{\rho}{2}\|\theta-z\|_{2}^{2}
\end{equation*}
\subsection{ADMM for consensus problems}
Consider the optimization problem given by
\begin{equation}\label{eq:gen_consensus}
\min_{\theta^{g}} \sum_{n=1}^{N} f_{n}(\theta^{g}),
\end{equation}
where $\theta^{g} \in \mathbb{R}^{n_{\theta}}$ and each term of the objective, i.e. $f_{n}: \mathbb{R}^{n_{\theta}} \rightarrow \mathbb{R} \cup \{+\infty \}$, is a proper, closed, convex function.\\
Suppose that $N$ processors are available to solve \eqref{eq:gen_consensus} and that, consequently, we are not interested in a centralized solution of the consensus problem. As explained in \cite{ADMMBoyd}, ADMM can be used to reformulate the problem so that each term of the cost function in \eqref{eq:gen_consensus} is handled by its own.\\
In particular, \eqref{eq:gen_consensus} can be reformulated as 
\begin{equation}\label{eq:rewritten_consensus}
\begin{aligned}
& \mbox{minimize } && \sum_{n=1}^{N} f_{n}(\theta_{n})\\
& \mbox{subject to } &&\theta_{n}-\theta^{g}=0\hspace{1cm} n=1,\ldots,N.
\end{aligned}
\end{equation}
Note that, thanks to the introduction of the consensus constraint, the cost function in \eqref{eq:rewritten_consensus} is now separable.\\
The augmented Lagrangian correspondent to \eqref{eq:rewritten_consensus} is given by
\begin{equation}\label{eq:lag_cons}
\mathcal{L}(\{\theta_{n}\}_{n=1}^{N},\theta^{g},\{\delta_{n}\}_{n=1}^{N})=\sum_{n=1}^{N}\left(f_{n}(\theta_{n})+\delta_{n}'(\theta_{n}-\theta^{g})+\frac{\rho}{2} \left\|\theta_{n}-\theta^{g}\right\|_{2}^{2}\right),
\end{equation}
and the ADMM iterations are
\begin{align}
&\theta_{n}^{(k+1)}=\underset{\theta_{n}}{\argmin} \hspace{0.15cm} \mathcal{L}_{n}(\theta_{n},\delta_{n}^{(k)},\theta^{g,(k)}), \ n=1,\ldots,N \label{eq:admm_cons:setp1}\\
&\theta^{g,(k+1)}=\frac{1}{N}\sum_{n=1}^{N} \left(\theta_{n}^{(k+1)}+\frac{1}{\rho}\delta_{n}^{(k)}\right), \label{eq:admm_cons:setp2}\\
&\delta_{n}^{(k+1)}=\delta_{n}^{(k)}+\rho\left(\theta_{n}^{(k+1)}-\theta^{g,(k+1)}\right), \ n=1,\ldots,N \label{eq:admm_cons:setp3}
\end{align}
with
\begin{equation*}
\mathcal{L}_{n}=f_{n}(\theta_{n})+(\delta_{n})'(\theta_{n}-\theta^{g})+\frac{\rho}{2}\|\theta_{n}-\theta^{g}\|_{2}^{2}.
\end{equation*}
Note that, on the one hand \eqref{eq:admm_cons:setp1} and \eqref{eq:admm_cons:setp3} can be carried out independently by each agent $n \in \{1,\ldots,N\}$, \eqref{eq:admm_cons:setp2} depends on all the updated local estimates. The global estimate should thus be updated in a \textquotedblleft fusion center\textquotedblright, where all the local estimates are collected and merged.

\section{Problem statement}\label{Sec:prob_for}
Assume that ($i$) measurements acquired by $N$ agents are available and that ($ii$) the behavior of the $N$ data-generating systems is described by the same known model. Suppose that some parameters of the model, $\theta_{n} \in \mathbb{R}^{n_{\theta}}$ with $n=1,\ldots,N$, are unknown and that their value has to to be retrieved from data. As the agents share the same model, it is also legitimate to assume that ($iii$) there exist a set of parameters $\theta^{g} \in \mathbb{R}^{n_{g}}$, with $n_{g} \leq n_{\theta}$, common to all the agents.\\
 
 We aim at ($i$) retrieving \emph{local} estimates of $\{\theta_{n}\}_{n=1}^{N}$, employing information available at the local level only, and ($ii$) identifying the \emph{global} parameter $\theta^{g}$ at the \textquotedblleft cloud\textquotedblright \ level, using the data collected from all the available sources. To accomplish these tasks ($i$) $N$ local processors and ($ii$)  and a \textquotedblleft cloud\textquotedblright, where the data are merged are needed.\\ 
 
 The considered estimation problem can be cast into a separable optimization problem, given by
 \begin{equation}\label{eq:problem}
 \begin{aligned}
 & \min_{\theta_{n}}  && \sum_{n=1}^{N} f_{n}(\theta_{n})\\
 & \mbox{s.t. } && F(\theta_{n})=\theta^{g},\\
 &&& \theta_{n} \in \mathcal{C}_{n}, \ n=1,\ldots,N
 \end{aligned}
 \end{equation}
 where $f_{n}: \mathbb{R}^{n_{\theta}} \rightarrow \mathbb{R}\cup\{+\infty\}$ is a closed, proper, convex function, $F: \mathbb{R}^{n_{\theta}} \rightarrow \mathbb{R}^{n_{g}}$ is a nonlinear operator and $\mathcal{C}_{n} \subset \mathbb{R}^{n_{\theta}}$ is a convex set representing constraints on the parameter values. Note that, constraints on the value of the global parameter can be enforced if  $\mathcal{C}_{n}=\mathcal{C}\cup\{\mathcal{C}_{n}\cap \breve{\mathcal{C}}\}$, with $\theta \in \mathcal{C}$.\\
Assume that the available data are the output/regressor pairs collected from each agent $n \in \{1,\ldots,N\}$ over an horizon of length $T \in \mathbb{N}$, i.e. $\{y_{n}(t),X_{n}(t)\}_{t=1}^{T}$. Relying on the hypothesis that the regressor/output relationship is well modelled as
\begin{equation}\label{eq:model_base}
 y_{n}(t)=X_{n}(t)'\theta_{n}+e_{n}(t),
 \end{equation}
 with $e_{n}(t) \in \mathbb{R}^{n_{y}}$ being a zero-mean additive noise independent of the regressor $X_{n}(t) \in \mathbb{R}^{n_{\theta}\times n_{y}}$, we will focus on developing a recursive algorithm to solve \eqref{eq:problem} with the local cost functions given by
\begin{equation}\label{eq:least_sqCost}
 f_{n}(\theta_{n})=\frac{1}{2}\sum_{t=1}^{T} \lambda_{n}^{T-t}\left\|y_{n}(t)-X_{n}(t)'\theta_{n}\right\|_{2}^{2}.
 \end{equation}
 The forgetting factor $\lambda_{n} \in (0,1]$ is introduced to be able to estimate time-varying parameters. Note that different forgetting factors can be chosen for different agents.

\begin{remark}{\textbf{ARX models}}\\
	Suppose that an AutoRegressive model with eXogenous inputs (ARX) has to be identified from data. The input/output relationship is thus given by
	\begin{align}\label{eq:arx_io}
	\nonumber y(t)&=\theta_{1}y(t-1)+\ldots+\theta_{n_{a}}y(t-n_{a})+\\
	&\hspace{1cm}+\theta_{n_{a}+1}u(t-n_{k}-1)+\ldots+\theta_{n_{a}+n_{b}}u(t-n_{k}-n_{b})+e(t)
	\end{align}
	where $u$ is the deterministic input, $\{n_{a}, n_{b}\}$ indicate the order of the system, $n_{k}$ is the input/output delay.\\
	Note that \eqref{eq:arx_io} can be recast as the output/regressor relationship with the regressor defined as
	\begin{equation}\label{eq:reg}
	X(t)=\begin{bmatrix}y(t-1)' & \ldots & y'(t-n_{a}) & u(t-n_{k}-1)' &  \ldots &  u(t-n_{k}-n_{b})'\end{bmatrix}'
	\end{equation}
	It is worth to point out that, in the considered framework, the parameters $n_{a}$, $n_{b}$ and $n_{k}$ are the same for all the N agents, as they are supposed to be described by the same model. \hfill $\blacksquare$
\end{remark}

\section{Collaborative estimation for full consensus}\label{Sec:1}
Suppose that the problem to be solve is \eqref{eq:gen_consensus}, i.e. we are aiming at achieving full consensus among $N$ agents. Consequently, the consensus constraint in \eqref{eq:problem} has to be modified as
\begin{equation*}
F(\theta_{n})=\theta^{g} \rightarrow \theta_{n}=\theta^{g}
\end{equation*}
and $\mathcal{C}_{n}=\mathbb{R}^{n_{\theta}}$, so that $\theta_{n} \in \mathcal{C}_{n}$ can be neglected for $n=1,\ldots,N$. Moreover, as we are focusing on the problem of collaborative least-squares estimation, we are interested in the particular case in which the local cost functions in \eqref{eq:rewritten_consensus} are equal to \eqref{eq:least_sqCost} .\\

Even if the considered problem can be solved in a centralized fashion, our goal is to obtain estimates of the unknown parameters both ($i$) at a local level and ($ii$) on the \textquotedblleft cloud\textquotedblright. With the objective of distributing the computation among the local processors and the \textquotedblleft cloud\textquotedblright, we propose $5$ approaches to address \eqref{eq:rewritten_consensus}.

\subsection{Greedy approaches}
All the proposed \textquoteleft greedy\textquoteright \ approaches rely on the use, by each local processor, of the standard Recursive Least-Squares (RLS) method (see~\cite{ljung1999system}) to update the local estimates, $\{\hat{\theta}_{n}\}_{n=1}^{N}$. Depending on the approach, $\{\hat{\theta}_{n}\}_{n=1}^{N}$ are then combined on the \textquotedblleft cloud\textquotedblright \ to update the estimate of the global parameter.\\

The first two methods that are used to compute the estimates of the unknown parameters both ($i$) locally and ($ii$) on the \textquotedblleft cloud\textquotedblright \ are:
\begin{enumerate}
	\item \textbf{Static RLS (S-RLS)} The estimate of the global parameter is computed as
	\begin{equation}\label{eq:mean}
	\hat{\theta}^{g}=\frac{1}{N} \sum_{n=1}^{N} \hat{\theta}_{n}(t).
	\end{equation}
	\item \textbf{Static Weighted RLS (SW-RLS)} Consider the matrices $\{\phi_{n}\}_{n=1}^{N}$, obtained applying standard RLS at each node (see~\cite{ljung1999system}), and assume that $\{\phi_{n}\}_{n=1}^{N}$ are always invertible. The estimate $\hat{\theta}^{g}$ is computed as the weighted average of the local estimates
	\begin{equation}\label{eq:w_mean}
	\hat{\theta}^{g}=\left(\sum_{n=1}^{N} \phi_{n}(t)^{-1}\right)^{-1}\left(\sum_{n=1}^{N} \phi_{n}(t)^{-1} \hat{\theta}_{n}(t)\right).
	\end{equation}
	 Considering that $\phi_{n}$ is an indicator of the accuracy of the $n$th local estimate, \eqref{eq:w_mean} allows to weight more the \textquotedblleft accurate \textquotedblright \ estimates then the \textquotedblleft inaccurate\textquotedblright \ ones.
\end{enumerate}
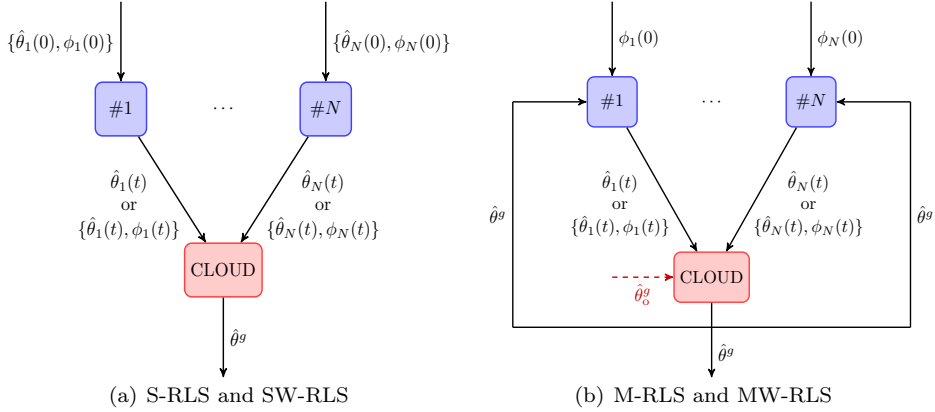
\begin{figure}[!tb]
	\centering
	\hspace{-1cm}
		\begin{tabular}[t]{cc}
			\subfigure[S-RLS and SW-RLS\label{Fig:SSchemes}]{
				\resizebox{6cm}{5cm}{\begin{tikzpicture}[node distance=1.3cm,>=stealth',bend angle=45,auto,->]
				\tikzstyle{agents}=[rectangle,rounded corners,thick,draw=blue!75,fill=blue!20,minimum size=1cm]
				\tikzstyle{dots}=[rectangle,fill=white]
				\tikzstyle{cloud}=[rectangle,rounded corners,thick,draw=red!75,fill=red!20,minimum size=1cm]
				\tikzstyle{aid}=[coordinate]
				
				\node[agents] (1) {$\# 1$};
				\node[dots,right of=1, node distance=2cm] (dots1) {$\boldmath{\cdots}$};
				\node[aid, above of=1, node distance=2cm] (aidIn1) {};
				\node[agents, right of = dots1, node distance=2cm](N) {$\# N$};
				\node[aid, above of=N, node distance=2cm] (aidInN) {};
				\node[cloud, below of= dots1, node distance=3cm] (C) {CLOUD};
				\node[aid, below of=C, node distance=2cm] (aidOut) {};
				\path (aidIn1) edge[thick]         node[swap] {$\{\hat{\theta}_{1}(0),\phi_{1}(0)\}$} (1)
				(aidInN) edge[thick]         node {$\{\hat{\theta}_{N}(0),\phi_{N}(0)\}$} (N)
				(1) edge[thick]         node[swap,yshift=0.5cm,xshift=0.5cm] {\begin{tabular}{c}$\hat{\theta}_{1}(t)$ \\
					or\\
					$\{\hat{\theta}_{1}(t),\phi_{1}(t)\}$\end{tabular}} (C)
				(N) edge[thick]         node[yshift=0.5cm,xshift=-0.5cm] {\begin{tabular}{c}$\hat{\theta}_{N}(t)$ \\
					or\\
					$\{\hat{\theta}_{N}(t),\phi_{N}(t)\}$\end{tabular}} (C)
				(C) edge[thick]         node {$\hat{\theta}^{g}$} (aidOut);				
				\end{tikzpicture}}
			} 
			\subfigure[M-RLS and MW-RLS \label{Fig:MSchemes}]{
				\resizebox{6cm}{5cm}{\begin{tikzpicture}[node distance=1.3cm,>=stealth',bend angle=45,auto,->]				
				\tikzstyle{agents}=[rectangle,rounded corners,thick,draw=blue!75,fill=blue!20,minimum size=1cm]
				\tikzstyle{dots}=[rectangle,fill=white]
				\tikzstyle{cloud}=[rectangle,rounded corners,thick,draw=red!75,fill=red!20,minimum size=1cm]
				\tikzstyle{aid}=[coordinate]
				
				\node[agents] (1) {$\# 1$};
				\node[dots,right of=1, node distance=2cm] (dots1) {$\boldmath{\cdots}$};
				\node[aid, above of=1, node distance=2cm] (aidIn1) {};
				\node[agents, right of = dots1, node distance=2cm](N) {$\# N$};
				\node[aid, above of=N, node distance=2cm] (aidInN) {};
				\node[cloud, below of= dots1, node distance=3.5cm] (C) {CLOUD};
				\node[aid, left of=C, node distance=2cm] (aidIn) {};
				\node[aid, below of=C, node distance=1cm] (aidOut1) {};
				\node[aid, right of=aidOut1, node distance=4cm] (aid1) {};
				\node[aid, right of=N, node distance=2cm] (aid2) {};
				\node[aid, left of=aidOut1, node distance=4cm] (aid3) {};
				\node[aid, left of=1, node distance=2cm] (aid4) {};
				\node[aid, below of=aidOut1, node distance=1cm] (aidOut2) {};
				\path (aidIn1) edge[thick]         node {$\phi_{1}(0)$} (1)
				(aidIn) edge[thick,dashed,black!20!red]         node[swap] {$\hat{\theta}_{\mathrm{o}}^{g}$} (C)
				(aidInN) edge[thick]         node {$\phi_{N}(0)$} (N)
				(1) edge[thick]         node [swap,yshift=0.5cm,xshift=0.5cm]{\begin{tabular}{c}$\hat{\theta}_{1}(t)$ \\
					or\\
					$\{\hat{\theta}_{1}(t),\phi_{1}(t)\}$\end{tabular}} (C)
				(N) edge[thick]         node[yshift=0.5cm,xshift=-0.5cm] {\begin{tabular}{c}$\hat{\theta}_{N}(t)$ \\
					or\\
					$\{\hat{\theta}_{N}(t),\phi_{N}(t)\}$\end{tabular}} (C)
				(C) edge[-,thick]         node {} (aidOut1)
				(aidOut1) edge[-,thick]         node {} (aid1)
				(aid1) edge[-,thick]         node[swap] {$\hat{\theta}^{g}$} (aid2)
				(aid2) edge[,thick]         node {} (N)
				(aidOut1) edge[-,thick]         node {} (aid3)
				(aid3) edge[-,thick]         node {$\hat{\theta}^{g}$} (aid4)
				(aid4) edge[,thick]         node {} (1)
				(aidOut1) edge[thick]         node {$\hat{\theta}^{g}$} (aidOut2);
				\end{tikzpicture}}}
		\end{tabular}
	\caption{Greedy approaches. Schematic of the information exchanges between the agents and the \textquotedblleft cloud\textquotedblright.}
	\label{Fig:SMschemes}
\end{figure}
S-RLS and SW-RLS allow to achieve our goal, i.e. ($i$) obtain a local estimate of the unknowns and ($ii$) compute $\hat{\theta}^{g}$ using all the information available. However, looking at the scheme in \figurename{~\ref{Fig:SSchemes}} and at Algorithm~\ref{algo1}, it can be noticed that the global estimate is not used at a local level.\\
\begin{algorithm}[!tb]
	\caption{S-RLS and SW-RLS}
	\label{algo1}
	~~\textbf{Input}: Sequence of observations $\{X_{n}(t),y_{n}(t)\}_{t=1}^T$, initial matrices $\phi_{n}(0) \in \mathbb{R}^{n_{\theta}\times n_{\theta}}$, initial estimates $\hat{\theta}_{n}(0) \in \mathbb{R}^{n_{\theta}}$,  $n=1,\ldots,N$
	\vspace*{.1cm}\hrule\vspace*{.1cm}
	\begin{enumerate}[label=\arabic*., ref=\theenumi{}]  
		\item \textbf{for} $t=1,\ldots,T$ \textbf{do}
		\begin{itemize}
			\item[] \hspace{-0.5cm} \textbf{\underline{Local}}
			\begin{enumerate}[label=\theenumi{}.\arabic*., ref=\theenumi{}.\theenumii{}]
				\item \textbf{for} $n=1,\ldots,N$ \textbf{do}
				\begin{enumerate}[label=\theenumii{}.\arabic*., ref=\theenumi{}.\theenumii{}.\theenumiii{}]
					\item \textbf{compute} $K_{n}(t)$, $\phi_{n}(t)$ and $\hat{\theta}_{n}(t)$ with standard RLS \cite{ljung1999system};
				\end{enumerate}
				\item \textbf{end for};
			\end{enumerate}
			\item[] \hspace{-0.5cm} \textbf{\underline{Global}}
			\begin{enumerate}[label=\theenumi{}.\arabic*., ref=\theenumi{}.\theenumii{}]
				\item \textbf{compute} $\hat{\theta}^{g}$;
			\end{enumerate}
		\end{itemize}
		\item \textbf{end}.
	\end{enumerate}
	\vspace*{.1cm}\hrule\vspace*{.1cm}
	~~\textbf{Output}: Local estimates $\{\hat{\theta}_{n}(t)\}_{t=1}^{T}$, $n=1,\ldots,N$, estimated global parameters $\{\hat{\theta}^{g}(t)\}_{t=1}^{T}$.
\end{algorithm}
Thanks to the dependence of $\hat{\theta}^{g}$ on all the available information, the local use of the global estimate might enhance the accuracy of $\{\hat{\theta}_{n}\}_{n=1}^{N}$. Motivated by this observation, we introduce two additional methods:
\begin{enumerate}
	\setcounter{enumi}{3}
	\item \textbf{Mixed RLS (M-RLS)}
	\item \textbf{Mixed Weighted RLS (MW-RLS)}
\end{enumerate}
While M-RLS relies on \eqref{eq:mean}, in MW-RLS the local estimates are combined as in \eqref{eq:w_mean}. However, as shown in \figurename{~\ref{Fig:MSchemes}} and outlined in Algorithm~\ref{algo2}, the global estimate $\hat{\theta}^{g}$ is fed to the each local processor and used to update the local estimates instead of their values at the previous step.
\begin{algorithm}[!tb]
	\caption{M-RLS and MW-RLS}
	\label{algo2}
	~~\textbf{Input}: Sequence of observations $\{X_{n}(t),y_{n}(t)\}_{t=1}^T$, initial matrices $\phi_{n}(0) \in \mathbb{R}^{n_{\theta} \times n_{\theta}}$, $n=1,\ldots,N$, initial estimate $\hat{\theta}_{\mathrm{o}}^{g}$.
	\vspace*{.1cm}\hrule\vspace*{.1cm}
	\begin{enumerate}[label=\arabic*., ref=\theenumi{}]  
		\item \textbf{for} $t=1,\ldots,T$ \textbf{do}
		\begin{itemize}
			\item[] \hspace{-0.5cm} \textbf{\underline{Local}}
			\begin{enumerate}[label=\theenumi{}.\arabic*., ref=\theenumi{}.\theenumii{}]
				\item \textbf{for} $n=1,\ldots,N$ \textbf{do}
				\begin{enumerate}[label=\theenumii{}.\arabic*., ref=\theenumi{}.\theenumii{}.\theenumiii{}]
					\item \textbf{set} $\hat{\theta}_{n}(t-1)=\hat{\theta}^{g}(t-1)$;
					\item \textbf{compute} $K_{n}(t)$, $\phi_{n}(t)$ and $\hat{\theta}_{n}(t)$ with standard RLS \cite{ljung1999system};
				\end{enumerate}
				\item \textbf{end for};
			\end{enumerate}
			\item[] \hspace{-0.5cm} \textbf{\underline{Global}}
			\begin{enumerate}[label=\theenumi{}.\arabic*., ref=\theenumi{}.\theenumii{}]
				\item \textbf{compute} $\hat{\theta}^{g}$;
			\end{enumerate}
		\end{itemize}
		\item \textbf{end}.
	\end{enumerate}
	\vspace*{.1cm}\hrule\vspace*{.1cm}
	~~\textbf{Output}: Local estimates $\{\hat{\theta}_{n}(t)\}_{t=1}^{T}$, $n=1,\ldots,N$, estimated global parameters $\{\hat{\theta}^{g}(t)\}_{t=1}^{T}$.
\end{algorithm}
Note that, especially at the beginning of the estimation horizon, the approximation made in M-RLS and MW-RLS might affect negatively some of the local estimates, e.g. the ones obtained by the agents characterized by a relatively small level of noise.
\begin{remark}
	While S-RLS and M-RLS require the local processors to transmit to the \textquotedblleft cloud\textquotedblright \ only $\{\hat{\theta}_{n}\}_{n=1}^{N}$, the pairs $\{\hat{\theta}_{n},\phi_{n}\}_{n=1}^{N}$ have to be communicated to the \textquotedblleft cloud\textquotedblright \ with both SW-RLS and MW-RLS (see~\eqref{eq:mean} and~\eqref{eq:w_mean}, respectively). Moreover, as shown in \figurename{~\ref{Fig:SMschemes}}, while S-RLS and SW-RLS require Node-to-Cloud-to-Node (N2C2N) transmissions, M-RLS and MW-RLS are based on a Node-to-Cloud (N2C) communication policy. \hfill $\blacksquare$
\end{remark}

\subsection{ADMM-based RLS (ADMM-RLS) for full consensus}\label{Subsec:ADMM-RLS_base}
Instead of resorting to greedy methods, we propose to solve \eqref{eq:gen_consensus} with ADMM.\\
Note that the same approach has been used to develop a fully distributed scheme for consensus-based estimation over Wireless Sensor Networks (WSNs) in \cite{Mateos2009}. However, our approach differs from the one introduced in \cite{Mateos2009} as we aim at exploiting the \textquotedblleft cloud\textquotedblright \ to attain consensus and, at the same time, we want local estimates to be computed by each node.\\
As the problem to be solved is equal to \eqref{eq:rewritten_consensus}, the ADMM iterations to be performed are \eqref{eq:admm_cons:setp1}-\eqref{eq:admm_cons:setp3}, i.e. 
\begin{align*}
&\hat{\theta}_{n}(T)^{(k+1)}=\underset{\theta_{n}}{\argmin} \  \left\{f_{n}(\theta_{n})+(\delta_{n}^{(k)})'(\theta_{n}-\hat{\theta}^{g,(k)})+\frac{\rho}{2}\|\theta_{n}-\hat{\theta}^{g,(k)}\|_{2}^{2}\right\},\\
&\hat{\theta}^{g,(k+1)}=\frac{1}{N}\sum_{n=1}^{N} \left(\theta_{n}^{(k+1)}+\frac{1}{\rho}\delta_{n}^{(k)}\right),\\
&\delta_{n}^{(k+1)}=\delta_{n}^{(k)}+\rho\left(\hat{\theta}_{n}^{(k+1)}(T)-\hat{\theta}^{g,(k+1)}\right), \ n=1,\ldots,N 
\end{align*}
with the cost functions $f_{n}$ defined as in \eqref{eq:lag_cons} and where the dependence on $T$ of the local estimates is stressed to underline that only the updates of $\hat{\theta}_{n}$ are directly influenced by the current measurements. Note that the update for $\hat{\theta}^{g}$ is a combination of the mean of the local estimates, i.e.~\eqref{eq:mean}, and the mean of the Lagrange multipliers.\\ 
As \eqref{eq:admm_cons:setp2}-\eqref{eq:admm_cons:setp3} are independent from the specific choice of $f_{n}(\theta_{n})$, we focus on the update of the local estimates, i.e.~\eqref{eq:admm_cons:setp1}, with the ultimate goal of finding recursive updates for $\hat{\theta}_{n}$.\\

Thanks to the characteristics of the chosen local cost functions, the closed-form solution for the problem in \eqref{eq:admm_cons:setp1} is given by
\begin{align}\label{eq:exp_sol2}
\hat{\theta}_{n}^{(k+1)}(T)&=\phi_{n}(T)\left(\mathcal{Y}_{n}(T)-\delta_{n}^{(k)}+\rho\hat{\theta}^{g,(k)}\right),\\
 \mathcal{Y}_{n}(t)&=\sum_{\tau=1}^{t}\lambda_{n}^{t-\tau}X_{n}(\tau)y_{n}(\tau), \ \ t=1,\ldots,T,\\
\phi_{n}(t)&=\left(\sum_{\tau=1}^{t}\lambda_{n}^{t-\tau}X_{n}(\tau)(X_{n}(\tau))'+\rho I_{n_{\theta}}\right)^{-1}, \ \ t=1,\ldots,T. \label{eq:phi_1}
\end{align} 
With the aim of obtaining recursive formulas to update $\hat{\theta}_{n}$, consider the local estimate obtained at $T-1$, which is given by
\begin{equation}\label{eq:prev_est2}
\hat{\theta}_{n}(T-1)=\phi_{n}(T-1)\left(\mathcal{Y}_{n}(T-1)+\rho\hat{\theta}^{g}(T-1)-\delta_{n}(T-1)\right),
\end{equation} 
with $\delta_{n}(T-1)$ and $\hat{\theta}^{g}(T-1)$ denoting the Lagrange multiplier and the global estimate computed at $T-1$, respectively. It has then to be proven that $\hat{\theta}_{n}^{(k)}(T-1)$ can be computed as a function of $\hat{\theta}(T-1)$, $y_{n}(T)$ and $X_{n}(T)$.\\

Consider the inverse matrix $\phi_{n}$ \eqref{eq:phi_1}, given by
\begin{align*}
&\phi_{n}(T)^{-1}=\mathcal{X}_{n}(T)+\rho I_{n_{\theta}},\\
&\mathcal{X}_{n}(t)=\sum_{\tau=1}^{t}\lambda_{n}^{t-\tau}X_{n}(\tau)(X_{n}(\tau))'.
\end{align*}
Based on \eqref{eq:phi_1}, it can be proven that $\phi_{n}(T)^{-1}$ can be computed as a function of $\phi_{n}(T-1)^{-1}$. In particular:
\begin{align}
\nonumber &\phi_{n}(T)^{-1}=\mathcal{X}_{n}(T)+\rho I_{n_{\theta}}=\\
\nonumber  &=\lambda_{n}\mathcal{X}_{n}(T-1)+X_{n}(T)(X_{n}(T))'+\rho I_{n_{\theta}}=\\
\nonumber &=\lambda_{n}\left[\mathcal{X}_{n}(T-1)+\rho I_{n_{\theta}}\right]+X_{n}(T)(X_{n}(T))'+(1-\lambda_{n})\rho I_{n_{\theta}}=\\
&=\lambda_{n}\phi_{n}(T-1)^{-1}+X_{n}(T)(X_{n}(T))'+(1-\lambda_{n})\rho I_{n_{\theta}}. \label{eq:phi_presimpl}
\end{align}
Introducing the extended regressor vector $\tilde{X}_{n}(T)$
\begin{equation}\label{eq:tildeX_1}
\tilde{X}_{n}(T)=\begin{bmatrix}
X_{n}(T) & \sqrt{(1-\lambda_{n})\rho}I_{n_{\theta}}
\end{bmatrix} \in \mathbb{R}^{n_{\theta}\times(n_{y}+n_{\theta})},
\end{equation}
\eqref{eq:phi_presimpl} can then be further simplified as
\begin{equation*}
\phi_{n}(T)^{-1}=\lambda_{n}\phi_{n}(T-1)^{-1}+\tilde{X}_{n}(T)(\tilde{X}_{n}(T))'.
\end{equation*}
Applying the matrix inversion lemma, the resulting recursive formulas to update $\phi_{n}$ are
\begin{align}
\mathcal{R}_{n}(T)&=\lambda_{n} I_{(n_{y}+n_{\theta})}+(\tilde{X}_{n}(T))'\phi_{n}(T-1)\tilde{X}_{n}(T), \label{eq:toinvert_1}\\
K_{n}(T)&=\phi_{n}(T-1)\tilde{X}_{n}(T)(\mathcal{R}_{n}(T))^{-1}, \label{eq:gain_2}\\
\phi_{n}(T)&=\lambda_{n}^{-1}\left(I_{n_{\theta}}-K_{n}(T)(\tilde{X}_{n}(T))'\right)\phi_{n}(T-1). \label{eq:phi_rec2}
\end{align} 
Note that the gain $K_{n}$ and matrix $\phi_{n}$ are updated as in standard RLS (see~\cite{ljung1999system}), with the exceptions of the increased dimension of the identity matrix in \eqref{eq:toinvert_1}  and the substitution of the regressor with $\tilde{X}_{n}$. Only when $\lambda_{n}=1$ the regressor $X_{n}$ and $\tilde{X}_{n}$ are equal. Moreover, observe that \eqref{eq:toinvert_1}-\eqref{eq:phi_rec2} are independent from $k$ and, consequently, $\{\mathcal{R}_{n},K_{n},\phi_{n}\}_{n=1}^{N}$ can be updated once fer step $t$.\\

Consider again \eqref{eq:exp_sol2}. Adding and subtracting
\begin{equation*}
\lambda_{n}\phi_{n}(T)\left[ \rho\hat{\theta}^{g}(T-1)-\delta_{n}(T-1)\right]
\end{equation*}
to \eqref{eq:exp_sol2}, the solution of \eqref{eq:admm_cons:setp1} corresponds to
\begin{align}
\nonumber \hat{\theta}_{n}^{(k+1)}(T)&=\phi_{n}(T)\left[\lambda_{n}\left(\mathcal{Y}_{n}(T-1)-\delta_{n}(T-1)+\rho \hat{\theta}^{g}(T-1)\right)+\right.\\
\nonumber & \hspace{-0.6cm}\left.+X_{n}(T)y_{n}(T)-\left(\delta_{n}^{(k)}-\lambda_{n}\delta_{n}(T-1)\right)+\rho\left(\hat{\theta}^{g,(k)}-\lambda_{n}\hat{\theta}^{g}(T-1)\right)\right]=\\
&\hspace{-0.5cm} =\hat{\theta}_{n}^{RLS}(T)+\hat{\theta}_{n}^{ADMM,(k+1)}(T), \label{eq:est_dec1}
\end{align}
with
\begin{align}
\nonumber&\hat{\theta}_{n}^{RLS}(T)=\phi_{n}(T)\left\{\lambda_{n}\left(\mathcal{Y}_{n}(T-1)+\rho\hat{\theta}^{g}(T-1)-\delta_{n}(T-1)\right)+\right.\\
&\hspace{2cm} \left.+X_{n}(T)y_{n}(T)\right\}, 	\label{eq:est_rls2} \\
&\hat{\theta}_{n}^{ADMM,(k+1)}(T)=\phi_{n}(T)\left[\rho\Delta_{g,\lambda_{n}}^{(k+1)}(T)-\Delta_{\lambda_{n}}^{(k+1)}(T)\right], \label{eq:est_admm2}
\end{align}
and
\begin{align}
\Delta_{g,\lambda_{n}}^{k+1}(T)&=\hat{\theta}^{g,(k)}-\lambda_{n}\hat{\theta}^{g}(T-1), \label{eq:admm_k+1_g}\\ \Delta_{\lambda_{n}}^{(k+1)}(T)&=\delta_{n}^{(k)}-\lambda_{n}\delta_{n}(T-1). \label{eq:admm_k+1_dual}
\end{align}
Observe that \eqref{eq:est_admm2} is independent from the past data-pairs $\{y_{n}(t),X_{n}(t)\}_{t=1}^{T}$, while \eqref{eq:est_rls2} depends on $\mathcal{Y}_{n}(T-1)$. Aiming at obtaining recursive formulas to update $\hat{\theta}_{n}$, the dependence of \eqref{eq:est_rls2} should be eliminated.\\

Consider \eqref{eq:est_rls2}. Exploiting \eqref{eq:phi_rec2} and \eqref{eq:prev_est2}, $\hat{\theta}_{n}^{RLS}(T)$ is given by
\begin{align}
\nonumber&\hat{\theta}_{n}^{RLS}(T)=\phi_{n}(T-1)\left\{ \left(\mathcal{Y}_{n}(T-1)+\rho\hat{\theta}^{g}(T-1)-\delta_{n}(T-1)\right)\right\}+\\
\nonumber &\hspace{0.5cm}-K_{n}(T)(\tilde{X}_{n}(T)')\phi_{n}(T-1)\left\{\left(\mathcal{Y}_{n}(T-1)+\rho\hat{\theta}^{g}(T-1)-\delta_{n}(T-1)\right)\right\}+\\
\nonumber &\hspace{0.5cm}+\phi_{n}(T)X_{n}(T)y_{n}(T)=\\ &\hspace{0.5cm}=\hat{\theta}_{n}(T-1)-K_{n}(T)(\tilde{X}_{n}(T))'\hat{\theta}_{n}(T-1)+\phi_{n}(T)X_{n}(T)y_{n}(T). \label{eq:rls_def1}
\end{align}
For \eqref{eq:rls_def1} to be dependent on the extended regressor only, we define the extended measurement vector
\begin{equation*}
\tilde{y}_{n}(T)=\begin{bmatrix}
(y_{n}(T))' & 0_{1\times n_{g}}
\end{bmatrix}'.
\end{equation*}
The introduction of $\tilde{y}_{n}$ yields \eqref{eq:rls_def1} can be modified as
\begin{equation*}
\hat{\theta}_{n}^{RLS}(T)=\hat{\theta}_{n}(T-1)-K_{n}(T)(\tilde{X}_{n}(T))'\hat{\theta}_{n}(T-1)+\phi_{n}(T)\tilde{X}_{n}(T)\tilde{y}_{n}(T).
\end{equation*}
Notice that the equality $\phi_{n}(T)\tilde{X}_{n}(T)=K_{n}(T)$ holds and it can be proven as follows
\begin{align*}
&\phi_{n}(T)\tilde{X}_{n}(T)=\lambda_{n}^{-1}\left(I_{n_{\theta}}-K_{n}(T)(\tilde{X}_{n}(T))'\right)\phi_{n}(T-1)\tilde{X}_{n}(T)=\\
&=\lambda_{n}^{-1}\left(I_{n_{\theta}}-\phi_{n}(T-1)\tilde{X}_{n}(T)(\mathcal{R}_{n}(T))^{-1}(\tilde{X}_{n}(T))'\right)\phi_{n}(T-1)\tilde{X}_{n}(T)=\\
&=\phi_{n}(T-1)\tilde{X}_{n}(T)\left(\lambda_{n}^{-1}I_{n_{\theta}}-\lambda_{n}^{-1}(\mathcal{R}_{n}(T))^{-1}(\tilde{X}_{n}(T))'\phi_{n}(T-1)\tilde{X}_{n}(T)\right)=\\
&=\phi_{n}(T-1)\tilde{X}_{n}(T)\left(\lambda_{n}^{-1}I_{n_{\theta}}+\right.\\
&\hspace{0.1cm}\left.-\lambda_{n}^{-1}(\lambda_{n} I_{(n_{y}+n_{\theta})}+(\tilde{X}_{n}(T))'\phi_{n}(T-1)\tilde{X}_{n}(T))^{-1}(\tilde{X}_{n}(T))'\phi_{n}(T-1)\tilde{X}_{n}(T)\right)=\\
&=\phi_{n}(T-1)\tilde{X}_{n}(T)\left(\lambda_{n}^{-1}I_{n_{\theta}}-\lambda_{n}^{-1}( I_{(n_{y}+n_{\theta})}+\lambda_{n}^{-1}(\tilde{X}_{n}(T))'\phi_{n}(T-1)\tilde{X}_{n}(T))^{-1}\hspace{-0.3cm}\cdot\right.\\
&\hspace{0.1cm} \left. \cdot (\tilde{X}_{n}(T))'\phi_{n}(T-1)\tilde{X}_{n}(T)\lambda_{n}^{-1}\right)=\\
&=\phi_{n}(T-1)\tilde{X}_{n}(T)\left(\lambda_{n}I_{n_{\theta}}+(\tilde{X}_{n}(T))'\phi_{n}(T-1)\tilde{X}_{n}(T)\right)^{-1}=K_{n}(T),
\end{align*}
where the matrix inversion lemma and \eqref{eq:gain_2}-\eqref{eq:phi_rec2} are used.\\
It can thus be proven that $\hat{\theta}_{n}^{RLS}$ can be updated as
\begin{equation}\label{eq:rls_full}
\hat{\theta}_{n}^{RLS}(T)=\hat{\theta}_{n}(T-1)+K_{n}(T)(\tilde{y}_{n}(T)-\tilde{X}_{n}(T)'\hat{\theta}_{n}(T-1)).
\end{equation}

While the update for $\hat{\theta}_{n}^{ADMM}$ \eqref{eq:est_admm2} depends on both the values of the Lagrange multipliers and the global estimates, $\hat{\theta}_{n}^{RLS}$ \eqref{eq:rls_full} is computed on the basis of the previous local estimate and the current measurements. Consequently, $\hat{\theta}_{n}^{RLS}$ is updated recursively.\\

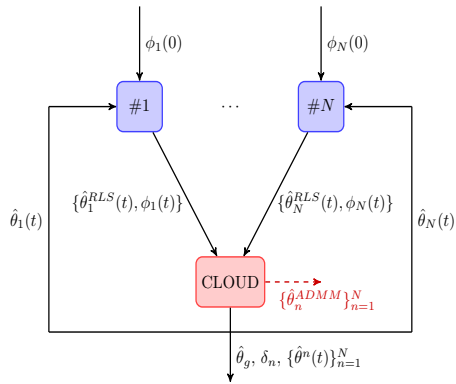
\begin{figure}[!tb]
 	\centering
 	\hspace{-1cm}
 			\resizebox{6cm}{5cm}{\begin{tikzpicture}[node distance=1.3cm,>=stealth',bend angle=45,auto,->]				
 				\tikzstyle{agents}=[rectangle,rounded corners,thick,draw=blue!75,fill=blue!20,minimum size=1cm]
 				\tikzstyle{dots}=[rectangle,fill=white]
 				\tikzstyle{cloud}=[rectangle,rounded corners,thick,draw=red!75,fill=red!20,minimum size=1cm]
 				\tikzstyle{aid}=[coordinate]
 				
 				\node[agents] (1) {$\# 1$};
 				\node[dots,right of=1, node distance=2cm] (dots1) {$\boldmath{\cdots}$};
 				\node[aid, above of=1, node distance=2cm] (aidIn1) {};
 				\node[agents, right of = dots1, node distance=2cm](N) {$\# N$};
 				\node[aid, above of=N, node distance=2cm] (aidInN) {};
 				\node[cloud, below of= dots1, node distance=3.5cm] (C) {CLOUD};
 				\node[aid, right of=C, node distance=2cm] (aidIn) {};
 				\node[aid, below of=C, node distance=1cm] (aidOut1) {};
 				\node[aid, right of=aidOut1, node distance=4cm] (aid1) {};
 				\node[aid, right of=N, node distance=2cm] (aid2) {};
 				\node[aid, left of=aidOut1, node distance=4cm] (aid3) {};
 				\node[aid, left of=1, node distance=2cm] (aid4) {};
 				\node[aid, below of=aidOut1, node distance=1cm] (aidOut2) {};
 				\path (aidIn1) edge[thick]         node {$\phi_{1}(0)$} (1)
 				(C) edge[thick,dashed,black!20!red]         node[swap] {\hspace{1.5cm}$\{\hat{\theta}_{n}^{ADMM}\}_{n=1}^{N}$} (aidIn)
 				(aidInN) edge[thick]         node {$\phi_{N}(0)$} (N)
 				(1) edge[thick]         node [swap,yshift=0.2cm,xshift=0.1cm]{
 					$\{\hat{\theta}_{1}^{RLS}(t),\phi_{1}(t)\}$} (C)
 				(N) edge[thick]         node[yshift=0.2cm,xshift=-0.1cm] {
 					$\{\hat{\theta}_{N}^{RLS}(t),\phi_{N}(t)\}$} (C)
 				(C) edge[-,thick]         node {} (aidOut1)
 				(aidOut1) edge[-,thick]         node {} (aid1)
 				(aid1) edge[-,thick]         node[swap] {$\hat{\theta}_{N}(t)$} (aid2)
 				(aid2) edge[,thick]         node {} (N)
 				(aidOut1) edge[-,thick]         node {} (aid3)
 				(aid3) edge[-,thick]         node {$\hat{\theta}_{1}(t)$} (aid4)
 				(aid4) edge[,thick]         node {} (1)
 				(aidOut1) edge[thick]         node {$\hat{\theta}_{g}$, $\delta_{n}$, $\{\hat{\theta}^{n}(t)\}_{n=1}^{N}$} (aidOut2);
 				\end{tikzpicture}}
 	\caption{ADMM-RLS. Schematic of the information exchanges between the agents and the \textquotedblleft cloud\textquotedblright when using a N2C2N communication scheme.}
 	\label{Fig:ADMMscheme}
 \end{figure}

Under the hypothesis that both $\hat{\theta}^{g}$ and $\delta_{n}$ are stored on the \textquotedblleft cloud\textquotedblright, it does seems legitimate to update $\hat{\theta}^{g}$ and $\delta_{n}$ on the \textquotedblleft cloud\textquotedblright, along with $\hat{\theta}_{n}^{ADMM}$. Instead, the partial estimates $\hat{\theta}_{n}^{RLS}$, $n=1,\ldots,N$, can be updated by the local processors. Thanks to this choice, the proposed method, summarized in Algorithm~\ref{algo3} and \figurename{~\ref{Fig:ADMMscheme}}, allows to obtain estimates both at the ($i$) agent and ($ii$) \textquotedblleft cloud\textquotedblright \ level.\\ 
Observe that, thanks to the independence of \eqref{eq:rls_full} from $k$, $\hat{\theta}_{n}^{RLS}$ can be updated once per step $t$. The local updates are thus regulated by a local clock and not by the one controlling the ADMM iterations on the \textquotedblleft cloud\textquotedblright.\\
Looking at \eqref{eq:toinvert_1}-\eqref{eq:phi_rec2} and \eqref{eq:rls_full}, it can be noticed that $\hat{\theta}_{n}^{RLS}$ is updated through standard RLS, with the exceptions that, at step $t \in \{1,\ldots,T\}$, the update depends on the previous local estimate $\hat{\theta}_{n}(t-1)$ instead of depending on $\hat{\theta}_{n}^{RLS}(t-1)$ and that the output/regressor pair $\{y_{n}(t),X_{n}(t)\}$ is replaced with $\{\tilde{y}_{n}(t),\tilde{X}_{n}(t)\}$. As a consequence, the proposed method can be easily integrated with pre-existing RLS estimators already available locally.
  
\begin{remark}
	Algorithm~\ref{algo1} requires the initialization of  the local and global estimates. If some data are available to be processed in a batch mode, $\hat{\theta}_{n}(0)$ can be chosen as the best linear model, i.e.
	\begin{equation*}
	\hat{\theta}_{n}(0)=\underset{\theta_{n}}{\argmin} \sum_{t=1}^{\tau} \|y_{n}(t)-X_{n}(t)'\theta\|_{2}^{2}
	\end{equation*}
	and $\hat{\theta}^{g}(0)$ can be computed as the mean of $\{P\hat{\theta}_{n}(0)\}_{n=1}^{N}$. Moreover, the matrices $\phi_{n}$, $n=1,\ldots,N$, can be initialized as $\phi_{n}(0)=\gamma I_{n_{\theta}}$, with $\gamma>0$.
	\hfill $\blacksquare$
\end{remark}

\begin{remark}
	The chosen implementation requires $\hat{\theta}_{n}^{RLS}$ and $\phi_{n}$ to be transmitted from the local processors to the \textquotedblleft cloud\textquotedblright \ at each step, while the \textquotedblleft cloud\textquotedblright has to communicate $\hat{\theta}_{n}$ to all the agents. As a consequence, the proposed approach is based on N2C2N transmissions. \hfill $\blacksquare$
\end{remark}

\begin{algorithm}[!tb]
	\caption{ADMM-RLS for full consensus (N2C2N)}
	\label{algo3}
	~~\textbf{Input}: Sequence of observations $\{X_{n}(t),y_{n}(t)\}_{t=1}^T$, initial matrices $\phi_{n}(0) \in \mathbb{R}^{n_{\theta} \times n_{\theta}}$, initial local estimates $\hat{\theta}_{n}(0)$, initial dual variables $\delta_{n,\mathrm{o}}$, $n=1,\ldots,N$, initial global estimate $\hat{\theta}_{\mathrm{o}}^{g}$, parameter $\rho \in \mathbb{R}^{+}$.
	\vspace*{.1cm}\hrule\vspace*{.1cm}
	\begin{enumerate}[label=\arabic*., ref=\theenumi{}]  
		\item \textbf{for} $t=1,\ldots,T$ \textbf{do}
		\begin{itemize}
			\item[] \hspace{-0.5cm} \textbf{\underline{Local}}
			\begin{enumerate}[label=\theenumi{}.\arabic*., ref=\theenumi{}.\theenumii{}]
				\item \textbf{for} $n=1,\ldots,N$ \textbf{do}
				\begin{enumerate}[label=\theenumii{}.\arabic*., ref=\theenumi{}.\theenumii{}.\theenumiii{}]
					\item \textbf{compute} $\tilde{X}_{n}(t)$ as in \eqref{eq:tildeX_1};
					\item \textbf{compute} $K_{n}(t)$ and $\phi_{n}(t)$ with \eqref{eq:gain_2}~-~\eqref{eq:phi_rec2};
					\item \textbf{compute} $\hat{\theta}_{n}^{RLS}(t)$ with \eqref{eq:rls_full};
				\end{enumerate}
			\item \textbf{end for};
			\end{enumerate}
			\item[] \hspace{-0.5cm} \textbf{\underline{Global}}
			\begin{enumerate}[label=\theenumi{}.\arabic*., ref=\theenumi{}.\theenumii{}]
				\item \textbf{do} 
				\begin{enumerate}[label=\theenumii{}.\arabic*., ref=\theenumi{}.\theenumii{}.\theenumiii{}]
				\item \textbf{compute} $\hat{\theta}_{n}^{ADMM,(k+1)}(t)$ with \eqref{eq:est_admm2}, $n=1,\ldots,N$;
				\item \textbf{compute} $\hat{\theta}^{g,(k+1)}(t)$ with \eqref{eq:admm_cons:setp2};
				\item \textbf{compute} $\delta_{n}^{(k+1)}$ with \eqref{eq:admm_cons:setp3}, $n=1,\ldots,N$;
			\end{enumerate}
		\item \textbf{until} a stopping criteria is satisfied (e.g. maximum number of iterations attained);
			\end{enumerate}
		\end{itemize}
		\item \textbf{end}.
	\end{enumerate}
	\vspace*{.1cm}\hrule\vspace*{.1cm}
	~~\textbf{Output}: Estimated global parameters $\{\hat{\theta}^{g}(t)\}_{t=1}^{T}$, estimated local parameters $\{\hat{\theta}_{n}(t)\}_{t=1}^{T}$, $n=1,\ldots,N$.
\end{algorithm}
\subsection{Example 1. Static parameters}
Suppose that $N$ data-generating systems are described by the following models
\begin{equation}\label{eq:syst1}
y_{n}(t)=0.9y_{n}(t-1)+0.4u_{n}(t-1)+e_{n}(t),
\end{equation}
where $y_{n}(t) \in \mathbb{R}$, $X_{n}(t)=\smallmat{y_{n}(t-1) & u_{n}(t-1)}'$, $u_{n}$ is known and is generated in this example as a sequence of i.i.d. elements uniformly distributed in the interval $\smallmat{2 & 3}$ and $e_{n} \sim \mathcal{N}(0,R_{n})$ is a white noise sequence, with  $\{R_{n} \in \mathbb{N}\}_{n=1}^{N}$ randomly chosen in the interval $\smallmat{1 & 30}$. Evaluating the effect of the noise on the output $y_{n}$ through the Signal-to-Noise Ratio $SNR_{n}$, i.e.
\begin{equation}\label{eq:snr}
\mathrm{SNR}_{n}=10\log{\frac{\sum_{t=1}^T\left(y_{n}(t)-e_{n}(t)\right)^2}{\sum_{t=1}^T e_{n}(t)^2}}~dB
\end{equation}
the chosen covariance matrices yield $\mbox{SNR}_{n}\in [7.8 \ 20.8]$~dB, $n=1,\ldots,N$. Note that \eqref{eq:syst1} can be equivalently written as
\begin{equation*}
y_{n}(t)=(X_{n}(t))'\theta^{g}+e_{n}(t) \mbox{ with } \theta^{g}=\smallmat{0.9 & 0.4}'
\end{equation*}
and the regressor $X_{n}(t)$ is defined as in \eqref{eq:reg}, i.e. $X_{n}=\smallmat{y_{n}(t-1) & u_{n}(t-1)}$.\\ 

Observe that the deterministic input sequences $\{u_{n}(t)\}_{t=1}^{T}$ are all different. However, they are all generated accordingly to the same distribution, as it seems reasonable to assume that systems described by the same model are characterized by similar inputs.\\

\begin{figure}[!tb]
	\centerline{
		\begin{tabular}[t]{cc}
			\subfigure[$\theta_{1}^{g}$ vs $\hat{\theta}_{1}^{g}$]{\includegraphics[scale=0.7]{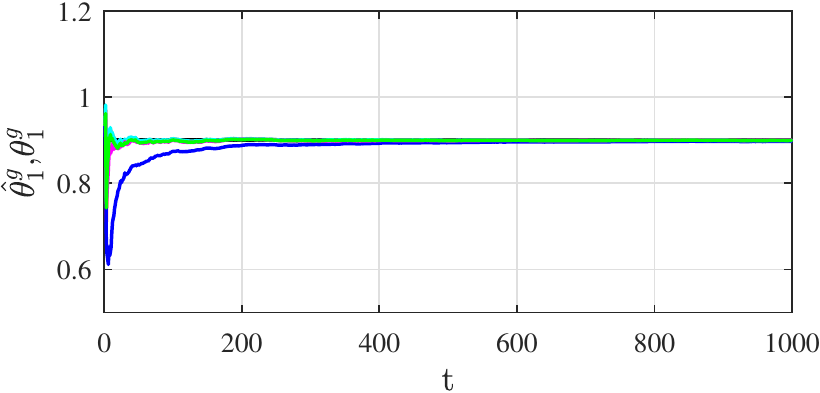}}
			\subfigure[$\theta_{2}^{g}$ vs $\hat{\theta}_{2}^{g}$]{\includegraphics[scale=0.7]{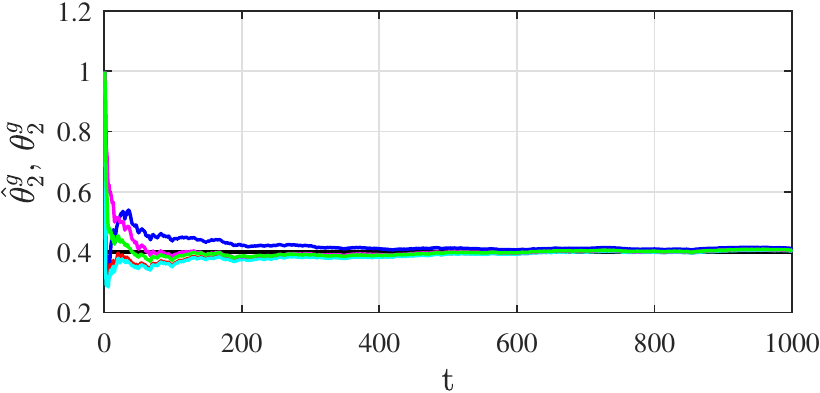}}\\
		\end{tabular}
	}
	\caption{Example 1. True vs estimated parameters. Black : true,
		red : C-RLS, blue : S-RLS, cyan : SW-RLS, magenta : M-RLS, green : MW-RLS.}
	\label{Fig:pramsGreedy1}
\end{figure}

Initializing $\phi_{n}$ as $\phi_{n}(0)=0.1I_{n_{\theta}}$, while $\hat{\theta}_{n}(0)$ and $\hat{\theta}_{\mathrm{o}}^{g}$ are sampled from the distributions $\mathcal{N}(\hat{\theta}^{g},2I_{n_{\theta}})$ and $\mathcal{N}(\hat{\theta}^{g},I_{n_{\theta}})$, respectively, and $\{\lambda_{n}=\Lambda\}_{n=1}^{N}$, with $\Lambda=1$, we first evaluate the performance of the greedy approaches. The actual parameter $\theta^{g}$ and the estimate obtained with the different greedy approaches are reported in \figurename{~\ref{Fig:pramsGreedy1}}.\\
Despite the slight difference performances in the first $300$ steps, which seems to be legitimate, the estimates obtained with SW-RLS, M-RLS and MW-RLS are similar. Moreover, $\hat{\theta}^{g}$ obtained with the different methods are comparable with respect with the estimate computed with C-RLS.\\
In particular, the similarities between the estimates obtained with M-RLS, MW-RLS and C-RLS prove that, in the considered case, the choice of the \textquotedblleft mixed\textquotedblright \ strategy allows to enhance the accuracy of $\hat{\theta}^{g}$. Comparing the estimates obtained with S-RLS and SW-RLS, observe that the convergence of the estimate to the actual value of $\theta^{g}$ tends to be faster if $\hat{\theta}^{g}$ is computed as in \eqref{eq:w_mean}.\\  

Setting $\rho=0.1$, the performance of the ADMM-RLS are assessed for different values of $N$ and $T$. Moreover, the retrieved estimates are  compared to the ones obtained with C-RLS and the greedy approaches.\\ 
The accuracy of the estimate $\hat{\theta}^{g}$ is assessed through the Root Mean Square Error (RMSE), i.e.
\begin{equation}\label{eq:rmse_glob}
\mathrm{RMSE}_{i}^{g}=\sqrt{\frac{\sum_{t=1}^T\left(\theta^{g}_{i}-\hat{\theta}_{i}^{g}(t)\right)^{2}}{T}}, \mbox{ } i=1,\ldots,n_{g}.
\end{equation}
\begin{table}[!tb]
	\begin{center}
		\caption{ADMM-RLS: $\|\mbox{RMSE}^{g}\|_{2}$} \label{Tab:RMSE1_1}
		{\footnotesize
			\begin{tabular}{|c|c|c|c|c|}
				\hline
				\backslashbox{\textbf{N}}{\textbf{T}}  & \textbf{10} & $\mathbf{10^{2}}$ & $\mathbf{10^{3}}$ & $\mathbf{10^{4}}$  \\
				\hline
				\textbf{2} & 1.07 & 0.33 & 0.16  & 0.10\\
				\hline
				\textbf{10} & 0.55 & 0.22 & 0.09  & 0.03 \\
				\hline
				$\mathbf{10^{2}}$ &  0.39 & 0.11 & 0.03 & 0.01 \\
				\hline			
		\end{tabular}} 
	\end{center}
\end{table}
As expected (see \tablename{~\ref{Tab:RMSE1_1}), the accuracy of the estimates tends to increase if the number of local processors $N$ and the estimation horizon $T$ increase. In the case $N=100$ and $T=1000$, the estimates obtained with both C-RLS and the SW-RLS and MW-RLS have comparable accuracy. See~\tablename{~\ref{Tab:RMSE_Comp1}}.\\
The estimates obtained with ADMM-RLS, C-RLS and MW-RLS are further compared in \figurename{~\ref{Fig:pramsAndErrors1}} and, as expected the three estimates are barely distinguishable.
\begin{table}[!tb]
	\begin{center}
		\caption{$\|\mbox{RMSE}^{g}\|_{2}$: C-RLS and greedy methods vs ADMM-RLS} \label{Tab:RMSE_Comp1}
		{\footnotesize
			\begin{tabular}{|c|c|c|c|c|c|c|}
				\cline{2-7}
				\multicolumn{1}{c|}{}  & \multicolumn{6}{c|}{Method}   \\
				\cline{2-7}
				\multicolumn{1}{c|}{}  & C-RLS &  S-RLS & SW-RLS & M-RLS & MW-RLS & ADMM-RLS  \\
				\hline
				$\|\mbox{RMSE}^{g}\|_{2}$ & 0.03  & 0.05 & 0.03 & 0.04 & 0.03 & 0.03 \\
				\hline
		\end{tabular}} 
	\end{center}
\end{table}
\begin{figure}[!tb]
	\centerline{
		\begin{tabular}[!tb]{cc}
			\subfigure[$\theta_{1}^{g}$ vs $\hat{\theta}_{1}^{g}$]{\includegraphics[scale=0.7]{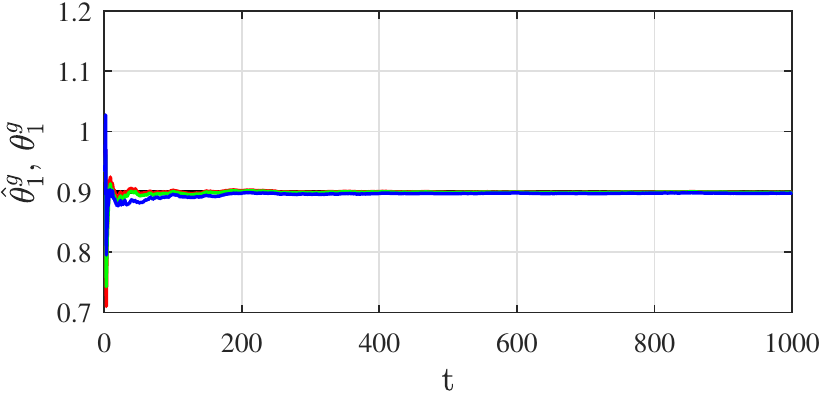}}
			\subfigure[$|\hat{\theta}_{1}^{g}-\theta_{1}^{g}|$]{\includegraphics[scale=0.7]{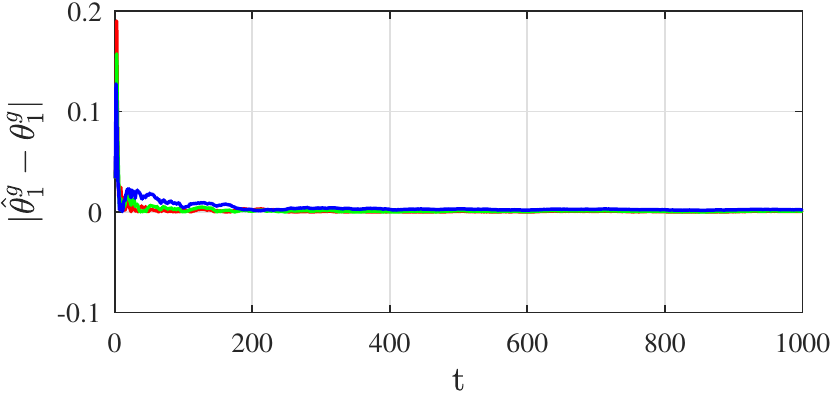}}\\
			\subfigure[$\theta_{2}^{g}$ vs $\hat{\theta}_{2}^{g}$]{\includegraphics[scale=0.7]{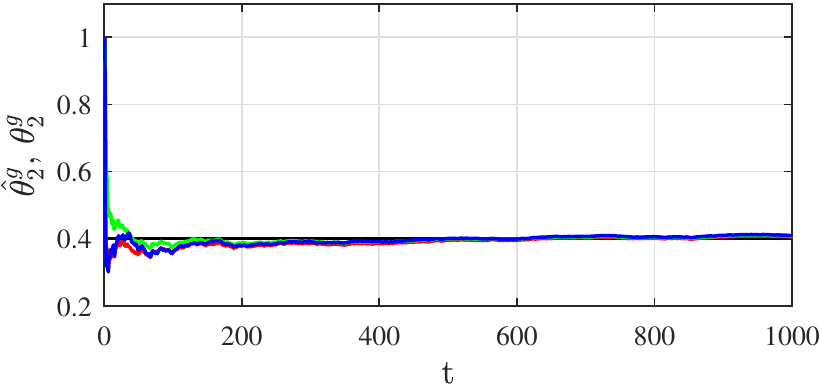}}
			\subfigure[$|\hat{\theta}_{2}^{g}-\theta_{2}^{g}|$]{\includegraphics[scale=0.7]{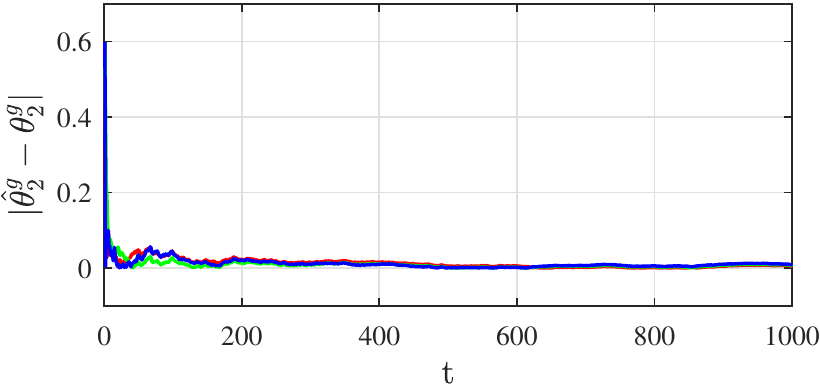}}\\			
		\end{tabular}
	}
	\caption{Example 1. Model parameters. Black : true,
		red : C-RLS, green : MW-RLS, blue : ADMM-RLS.}
	\label{Fig:pramsAndErrors1}
\end{figure}
 Thus the proposed ADMM-RLS algorithm, which uses local estimates and the cloud, is able to obtain good accuracy versus the fully centralized approach. Moreover, ADMM-RLS allows to retrieve estimates as accurate as the ones obtained with the MW-RLS, i.e. the greedy approach associated with the least RMSE.\\
 \subsubsection{Non-informative agents}
 Using the previously introduced initial setting and parameters, lets assume that some of the available data sources are non-informative, i.e. some systems are not excited enough to be able to retrieve locally an accurate estimate of all the unknown parameters \cite{ljung1999system}. Null input sequences and white noise sequences characterized by $R_{n}=10^{-8}$ are used to simulate the behavior of the $N_{ni}\leq N$ non-informative agents.\\
 
Consider the case $N=100$ and $T=5000$. The performance of ADMM-RLS are studied under the hypothesis that an increasing number $N_{ni}$ of systems is non-informative. Looking at the RMSEs in \tablename{~\ref{Tab:RMSE_Nni1}} and the estimates reported in \figurename{~\ref{Fig:pramsvsNni}}, it can be noticed that the quality of the estimate starts to decrease only when half of the available systems are non-informative.
 \begin{table}[!tb]
	\begin{center}
		\caption{Example 1. $\|\mbox{RMSE}^{g}\|_{2}$ vs $N_{ni}$} \label{Tab:RMSE_Nni1}
		{\footnotesize
			\begin{tabular}{|c|c|c|c|c|}
				\cline{2-5}
				\multicolumn{1}{c|}{}  & \multicolumn{4}{c|}{$N_{ni}$}   \\
				\cline{2-5}
				\multicolumn{1}{c|}{}  & 1 & 10 & 20 & 50 \\
				\hline
				$\|\mbox{RMSE}^{g}\|_{2}$ & 0.02 & 0.02 & 0.02 & 0.03 \\
				\hline		
		\end{tabular}} \vspace{-0.5cm}
	\end{center}
\end{table}
In case of $N_{ni}=20$, the estimates obtained with ADMM-RLS are then compared with the ones computed with C-RLS and the greedy approaches. As it can be noticed from the RMSEs reported in \tablename{~\ref{Tab:RMSE_Noninf1}}, in presence of non-informative agents SW-RLS tends to perform better than the other greedy approaches and the accuracy of the estimates obtained with C-RLS, SW-RLS and ADMM-RLS are comparable.
\begin{table}[!tb]
 	\begin{center}
 		\caption{Example 1. $\|\mbox{RMSE}^{g}\|_{2}$: $20\%$ of non-informative agents} \label{Tab:RMSE_Noninf1}
 		{\footnotesize
 			\begin{tabular}{|c|c|c|c|c|c|c|}
 				\cline{2-7}
 				\multicolumn{1}{c|}{}  & \multicolumn{6}{c|}{Method}   \\
 				\cline{2-7}
 				\multicolumn{1}{c|}{}  & C-RLS & S-RLS & SW-RLS & M-RLS & MW-RLS & ADMM-RLS  \\
 				\hline
 				$\|\mbox{RMSE}^{g}\|_{2}$ & 0.02 & 0.03 & 0.02 & 0.07 & 0.03 & 0.02 \\
 				\hline
 		\end{tabular}} \vspace{-0.5cm}
 	\end{center}
 \end{table}
 \begin{figure}[!tb]
 	\centerline{
 		\begin{tabular}[t]{cc}
 			\subfigure[$\theta_{1}^{g}$ vs $\hat{\theta}_{1}^{g}$]{\includegraphics[scale=0.7]{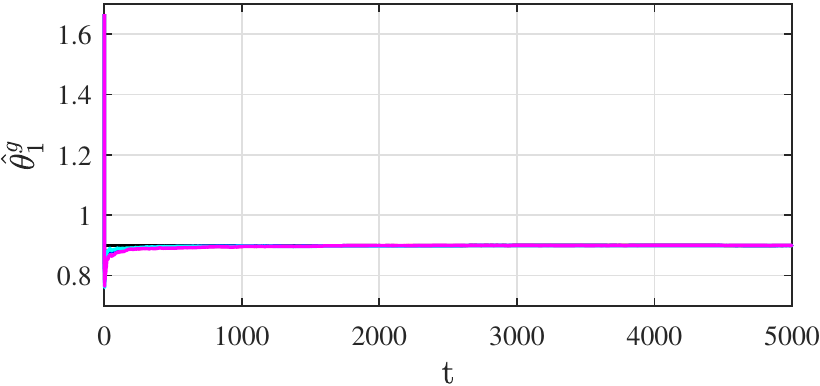}}
 			\subfigure[$\theta_{2}^{g}$ vs $\hat{\theta}_{2}^{g}$]{\includegraphics[scale=0.7]{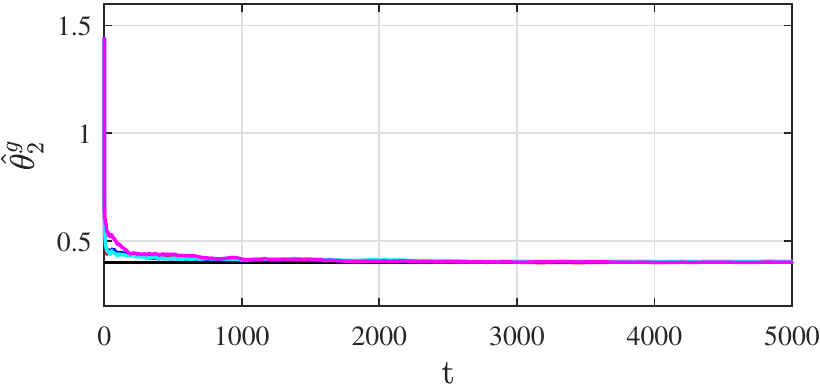}}		
 		\end{tabular}
 	}
 	\caption{Example 1. Model parameters vs $N_{ni}$. Black : true,
 		red : $N_{ni}=1$, blue : $N_{ni}=10$, cyan : $N_{ni}=20$, magenta : $N_{ni}=50$.}
 	\label{Fig:pramsvsNni}
 \end{figure}
 \subsubsection{Agents failure}
 Consider again $N=100$ and $T=5000$ and suppose that, due to a change in the behavior of $N_{f}$ local agents the parameters of their models suddenly assume different values with respect to $\smallmat{0.9 & 0.4}$. We study the performance of ADMM-RLS under the hypothesis that the change in the value of the parameters happens at an unknown instant $t_{n}$, randomly chosen in the interval $[1875,3750]$~samples, and simulating the change in the local parameters using  $\theta_{n,1}$ sampled from the distribution $\mathcal{U}_{\smallmat{0.2 & 0.21}}$ and $\theta_{n,2}$ sampled from $\mathcal{U}_{\smallmat{1.4 & 1.43}}$ after $t_{n}$.\\
 Observe that it might be beneficial to use a non-unitary forgetting factor, due to the change in the local parameters. Consequently, $\lambda_{n}$, $n=1,\ldots,N$, is set to $0.99$ for all the $N$ agents.\\
 The performance of ADMM-RLS are initially assessed considering an increasing number of systems subject to failure. See \tablename{~\ref{Tab:RMSE2_Nf}} and \figurename{~\ref{Fig:pramsAndErrors2FailADMM}}.
 \begin{table}[!tb]
 	\begin{center}
 		\caption{Example 1. ADMM-RLS: $\|\mbox{RMSE}^{g}\|_{2}$ vs $N_f$} \label{Tab:RMSE2_Nf}
 		{\footnotesize
 			\begin{tabular}{|c|c|c|c|c|}
 				\cline{2-5}
 				\multicolumn{1}{c|}{}  & \multicolumn{4}{c|}{$N_{f}$}   \\
 				\cline{2-5}
 				\multicolumn{1}{c|}{}  & 1 & 10 & 20 & 50  \\
 				\hline
 				$\|\mbox{RMSE}^{g}\|_{2}$ & 0.03 & 0.03 & 0.03 & 0.04\\
 				\hline		
 		\end{tabular}} \vspace{-0.5cm}
 	\end{center}
 \end{table}
\begin{figure}[!tb]
	\centerline{
		\begin{tabular}[!tb]{cc}
			\subfigure[$\theta_{1}^{g}$ vs $\hat{\theta}_{1}^{g}$]{\includegraphics[scale=0.7]{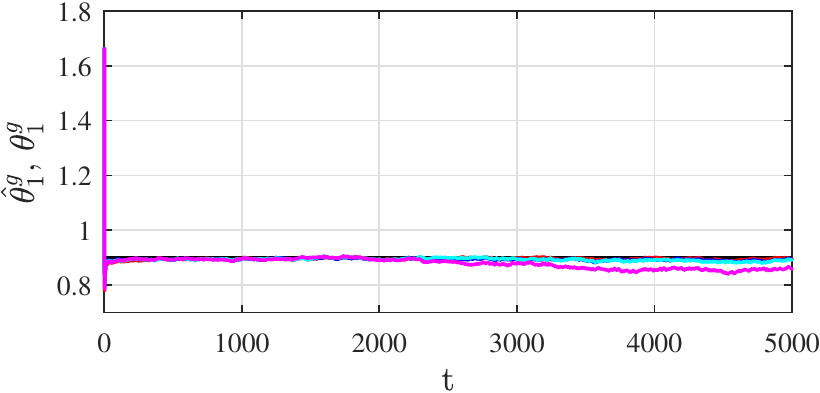}}
			\subfigure[$|\hat{\theta}_{1}^{g}-\theta_{1}^{g}|$]{\includegraphics[scale=0.7]{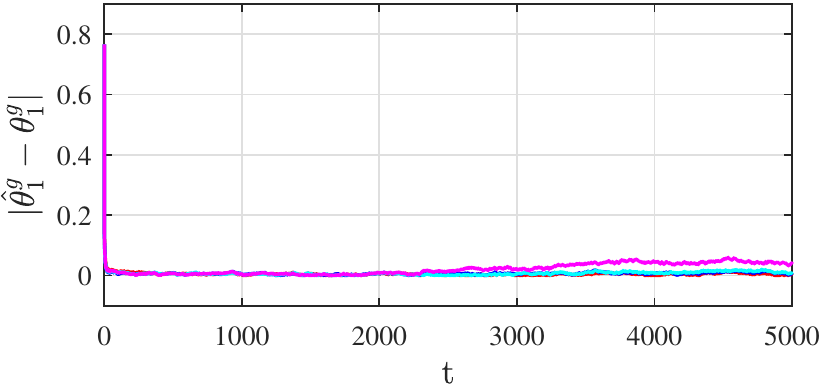}}\\
			\subfigure[$\theta_{2}^{g}$ vs $\hat{\theta}_{2}^{g}$]{\includegraphics[scale=0.7]{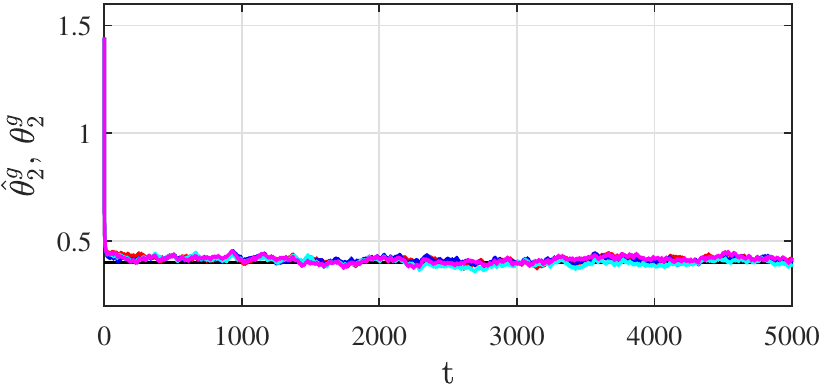}}
			\subfigure[$|\hat{\theta}_{2}^{g}-\theta_{2}^{g}|$]{\includegraphics[scale=0.7]{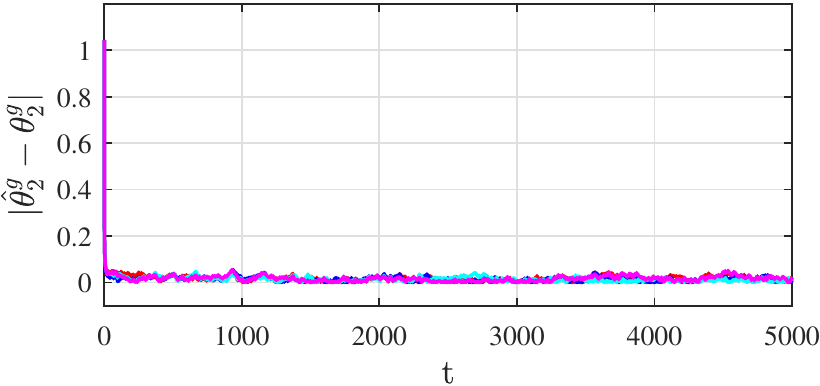}}\\
		\end{tabular}
	}
	\caption{Example 1. Model parameters vs $N_{f}$. Black : true,
		red : $N_{f}=1$, blue : $N_{f}=10$, cyan : $N_{f}=20$, magenta : $N_{f}=50$.}
	\label{Fig:pramsAndErrors2FailADMM}
\end{figure}
Observe that the failure of the agents seems not to influence the accuracy of the obtained estimates if $N_{f}\neq 50$. The use of ADMM-RLS thus allows to compute accurate global estimates even when some of the agent experience a failure.
\subsection{Example 2. Time-varying parameters}
The presence of the forgetting factor in the cost functions $f_{n}$ (see~\eqref{eq:least_sqCost}) allows to estimate time-varying parameters, as it enables to weight differently past and currently collected data.\\
Suppose that the behavior of $N$ systems is described by the ARX model
\begin{equation}\label{eq:syst2}
y_{n}(t+1)=\theta_{1}^{g}(t)y_{n}(t-1)+\theta_{2}^{g}(t)u_{n}(t-1)+e_{n}(t)
\end{equation}
where $\theta_{1}^{g}=0.9\sin{(x)}$ and $\theta_{2}^{g}=0.4\cos{(x)}$, with $x \in [0,2\pi]$, and $u_{n} \sim \mathcal{U}_{\smallmat{2 & 3}}$. The white noise sequences $e_{n}\sim \mathcal{N}(0,R_{n})$, $n=1,\ldots,N$, have covariances $R_{n}$ randomly selected in the interval $\smallmat{ 1 & 30 }$ yielding to $SNR_{n} \in \smallmat{ 2.4 & 6.5 }~\mbox{dB}$.\\
Considering an estimation horizon $T=1000$, imposing  $\phi_{n}$ as $\phi_{n}(0)=0.1I_{n_{\theta}}$, while $\hat{\theta}_{n}(0)$ and $\hat{\theta}_{\mathrm{o}}^{g}$ are sampled from the distributions $\mathcal{N}(\hat{\theta}^{g},2I_{n_{\theta}})$ and $\mathcal{N}(\hat{\theta}^{g},I_{n_{\theta}})$, respectively, $\rho=0.1$ and setting $\{\lambda_{n}=\Lambda\}_{n=1}^{N}$, with $\Lambda=0.95$, the performances of ADMM-RLS are compared with the ones of C-RLS and the four greedy approaches. See \tablename{~\ref{Tab:RMSE_Comp2}}. As for the case where time-invariant parameters have to be estimated (see Example~1), SW-RLS and MW-RLS tend to perform slightly better than the other greedy approaches. Note that the accuracy of the estimates C-RLS, SW-RLS and MW-RLS is comparable.\\
\begin{table}[!tb]
	\begin{center}
		\caption{Example 2. $\|\mbox{RMSE}^{g}\|_{2}$ vs Method} \label{Tab:RMSE_Comp2}
		{\footnotesize
			\begin{tabular}{|c|c|c|c|c|c|c|}
				\cline{2-7}
				\multicolumn{1}{c|}{}  & \multicolumn{6}{c|}{Method}   \\
				\cline{2-7}
				\multicolumn{1}{c|}{}  & C-RLS & S-RLS & SW-RLS & M-RLS & MW-RLS & ADMM-RLS  \\
				\hline
				$\|\mbox{RMSE}^{g}\|_{2}$ & 0.08 & 0.10 & 0.08 & 0.09 & 0.08 & 0.08 \\
				\hline		
		\end{tabular}} \vspace{-0.5cm}
	\end{center}
\end{table}
\begin{figure}[!tb]
	\centerline{
		\begin{tabular}[t]{cc}
			\subfigure[$\theta_{1}^{g}$ vs $\hat{\theta}_{1}^{g}$]{\includegraphics[scale=0.7]{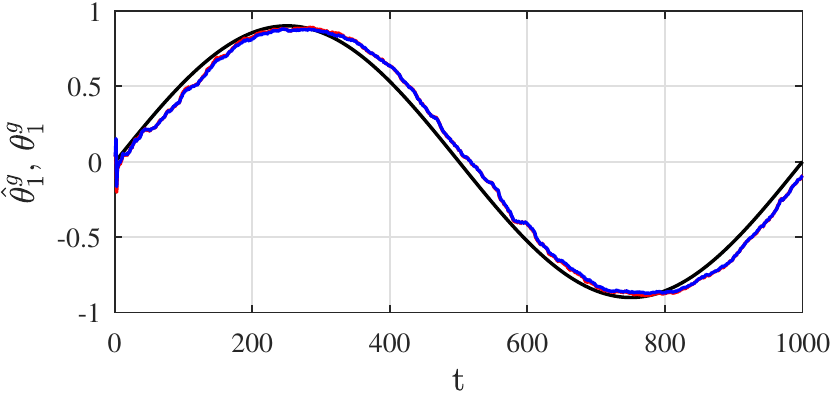}}
			\subfigure[$|\hat{\theta}_{1}^{g}-\theta_{1}^{g}|$]{\includegraphics[scale=0.7]{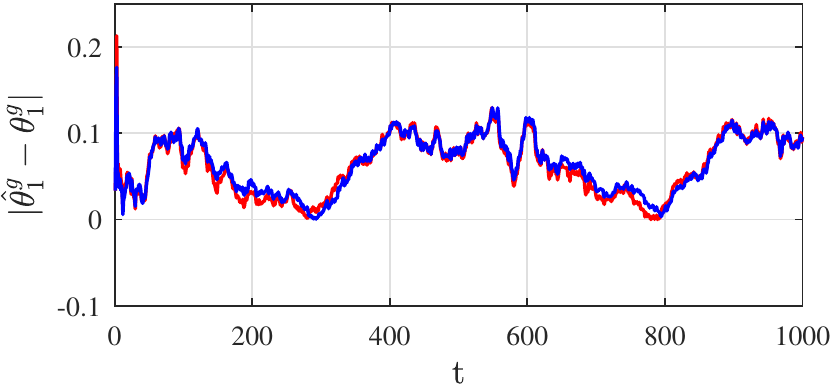}}\\
			\subfigure[$\theta_{2}^{g}$ vs $\hat{\theta}_{2}^{g}$]{\includegraphics[scale=0.7]{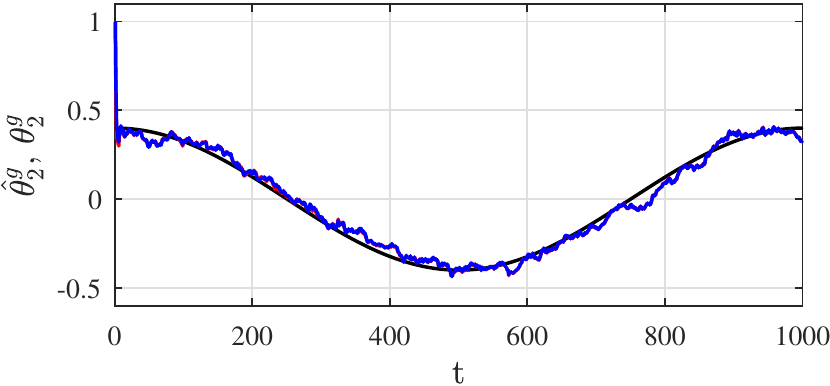}}
			\subfigure[$|\hat{\theta}_{2}^{g}-\theta_{2}^{g}|$]{\includegraphics[scale=0.7]{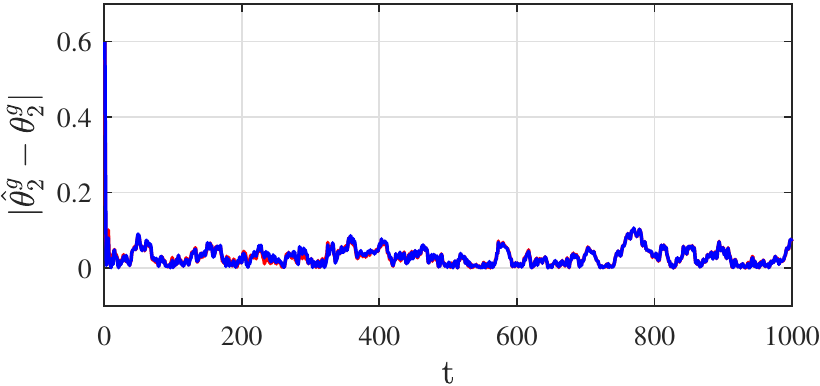}}\\			
		\end{tabular}
	}
	\caption{Example 2. True vs estimated model parameters. Black : true,
		red : C-RLS, blue : ADMM-RLS.}
	\label{Fig:pramsAndErrors2}
\end{figure}
\figurename{~\ref{Fig:pramsAndErrors2}} reports the actual global parameters and the estimates obtained with C-RLS and ADMM-RLS, along with the estimation errors. As already observed, the accuracy of the estimates computed with C-RLS and ADMM-RLS is comparable. 
\section{Collaborative estimation for partial consensus}\label{Sec:2}
Consider the more general hypothesis that there exist a parameter vector $\theta^g \in \mathbb{R}^{n_{g}}$, with $n_{g}\leq n_{\theta}$ such that:
\begin{equation}\label{eq:partial_constr}
P\theta_{n}=\theta^{g} \ \ \forall n \in \{1,\ldots,N\},
\end{equation}
where $P \in \mathbb{R}^{n_{g}\times n_{\theta}}$ is a matrix assumed to be known a priori. The problem that we want to solve is then given by
\begin{equation}\label{eq:prob_locglob}
\begin{aligned}
&\min_{\{\theta_{n}\}_{n=1}^{N}}&& \sum_{n=1}^{N} f_{n}(\theta_{n})\\
& \mbox{s.t. } && P\theta_{n}=\theta^{g}, \ \ n=1,\ldots,N,
\end{aligned}
\end{equation}
with $f_{n}$ defined as in \eqref{eq:least_sqCost}. Note that \eqref{eq:prob_locglob} corresponds to \eqref{eq:problem} with the consensus constraint modified as
\begin{equation*}
F(\theta_{n})=\theta^{g} \rightarrow P\theta_{n}=\theta^{g}.
\end{equation*}
The considered consensus constraint allows to enforce consensus over a linear combination of the components of $\theta_{n}$. Note that, through proper choices of $P$, different settings can be considered, e.g. if $P=I_{n_{\theta}}$ then $\theta_{n}=\theta^{g}$ and thus \eqref{eq:prob_locglob} is equal to \eqref{eq:gen_consensus}. We can also enforce consensus only over some components of $\theta_{n}$, so that some of the unknowns are assumed to be \emph{global} while others are supposed to assume a different value for each agent.\\
As we are interested in obtaining an estimate for both $\{\theta_{n}\}_{n=1}^{N}$ and $\theta^{g}$, note that \eqref{eq:prob_locglob} cannot be solved resorting to a strategy similar to C-RLS (see Appendix~\ref{Appendix:A}). In particular, even if properly modified, a method as C-RLS would allow to compute an estimate for the global parameter only.\\

The ADMM iterations to solve problem \eqref{eq:prob_locglob} are given by
\begin{align}
&\hat{\theta}_{n}^{(k+1)}(T)=\underset{\theta_{n}}{\argmin} \ \mathcal{L}(\theta_{n},\hat{\theta}^{g,(k)},\delta_{n}^{(k)}),
\label{step:loc_est}\\
&\hat{\theta}^{g,(k+1)}=\underset{\theta^{g}}{\argmin} \ \mathcal{L}(\{\hat{\theta}_{n}^{(k+1)}(T)\}_{n=1}^{N},\theta^{g},\{\delta_{n}^{(k)}\}_{n=1}^{N}), \label{step:glob_est}\\
&\delta_{n}^{(k+1)}=\delta_{n}^{(k)}+\rho(P\hat{\theta}_{n}^{(k+1)}(T)-\hat{\theta}^{g,(k+1)}), \label{step:dual_est}
\end{align}
with $k \in \mathbb{N}$ indicating the ADMM iteration, $\rho \in \mathbb{R}^{+}$ being a tunable parameter, $\delta_{n} \in \mathbb{R}^{n_{g}}$ representing the Lagrange multiplier and the augmented Lagrangian $\mathcal{L}$ given by
	\begin{equation}
\mathcal{L}=\sum_{n=1}^{N}\left\{f_{n}(\theta_{n})+\delta_{n}'(P\theta_{n}-\theta^{g})+\frac{\rho}{2}\left\|P\theta_{n}-\theta^{g}\right\|_{2}^{2}\right\}.
\end{equation} 
Note that the dependence on $T$ is explicitly indicated only for the local estimates $\hat{\theta}_{n}$, as they are the only quantities directly affected by the measurement and the regressor at $T$.\\
Consider the update of the estimate $\hat{\theta}^{g}$. The closed form solution for \eqref{step:glob_est} is
\begin{equation}\label{eq:glob_est}
\hat{\theta}^{g,(k+1)}=\frac{1}{N}\sum_{n=1}^{N}\left(P\hat{\theta}_{n}^{(k+1)}(T)+\frac{1}{\rho}\delta_{n}^{(k)}\right).
\end{equation}
The estimate of the global parameter is thus updated through the combination of the mean of $\{\delta_{n}\}_{n=1}^{N}$ and the mean of $\{P\hat{\theta}_{n}^{(k+1)}(T)\}_{n=1}^{N}$. As expected, \eqref{eq:glob_est} resembles \eqref{eq:admm_cons:setp2}, where the local estimates are replaced by a linear combination of their components.\\
Consider the update for the estimate of the local parameters. The close form solution for \eqref{step:loc_est} is given by:
\begin{align}
\hat{\theta}_{n}^{(k+1)}(T)&=\phi_{n}(T)\left\{\mathcal{Y}_{n}(T)+P'(\rho \hat{\theta}^{g,(k)}-\delta_{n}^{(k)}) \right\}, \label{eq:est_P}\\
\mathcal{Y}_{n}(t)&=\sum_{\tau=1}^{t} \lambda_{n}^{t-\tau}X_{n}(\tau)y_{n}(\tau), \ \ t=1,\ldots,T,\\
\phi_{n}(t)&=\left(\left[\sum_{\tau=1}^{t}\lambda_{n}^{t-\tau}X_{n}(\tau)X_{n}(\tau)' \right] +\rho P'P\right)^{-1}, \ \ t=1,\ldots,T. \label{eq:phi_form1}
\end{align}
As also in this case we are interested in obtaining recursive formulas for the local updates, consider $\hat{\theta}_{n}(T-1)$, defined as
\begin{equation}
\hat{\theta}_{n}(T-1)=\phi_{n}(T-1)\left(\mathcal{Y}_{n}(T-1)+P'(\rho\hat{\theta}^{g}(T-1)-\delta_{n}(T-1)) \right),
\end{equation}
where $\phi_{n}(T-1)$ is equal to \eqref{eq:phi_form1}, and $\hat{\theta}^{g}(T-1)$ and $\delta_{n}(T-1)$ are the global estimate and the Lagrange multiplier obtained at $T-1$, respectively.\\
Observe that the following equalities hold
\begin{align*}
\phi_{n}(T)&=\left(\mathcal{X}_{n}(T)+\rho P'P\right)^{-1}=\\
&=\left(\lambda_{n}\mathcal{X}_{n}(T-1)+X_{n}(T)X_{n}(T)'+\rho P'P\right)^{-1}=\\
&=\left(\lambda_{n}\left(\mathcal{X}_{n}(T-1)+\rho P'P\right)+X_{n}(T)X_{n}(T)'+\rho(1-\lambda_{n}) P'P\right)^{-1}=\\
&=\left(\lambda_{n}\phi_{n}(T-1)^{-1}+X_{n}(T)X_{n}(T)'+\rho(1-\lambda_{n}) P'P\right)^{-1},
\end{align*}
with
\begin{equation*}
\mathcal{X}_{n}(t)=\sum_{\tau=1}^{t}\lambda_{n}^{t-\tau}X_{n}(\tau)(X_{n}(\tau))', \ \ t=1,\ldots,T.
\end{equation*}
Introducing the extended regressor
\begin{equation}\label{eq:tildeX_2}
\tilde{X}_{n}(T)=\begin{bmatrix} X_{n}(T) & \sqrt{\rho(1-\lambda_{n})}P' \end{bmatrix} \in \mathbb{R}^{n_{\theta}\times(n_{y}+n_{g})}
\end{equation}
and applying the matrix inversion lemma, it can be proven that $\phi_{n}$ can be updated as
\begin{align}
\mathcal{R}_{n}(T)&=\lambda_{n} I_{(n_{y}+n_{g})}+(\tilde{X}_{n}(T))'\phi_{n}(T-1)\tilde{X}_{n}(T), \label{eq:toinvert_2}\\
K_{n}(T)&=\phi_{n}(T-1)\tilde{X}_{n}(T)\left(\mathcal{R}_{n}(T)\right)^{-1}, \label{eq:gain_3}\\
\phi_{n}(T)&=\lambda_{n}^{-1}(I_{n_{\theta}}-K_{n}(T)(\tilde{X}_{n}(T))')\phi_{n}(T-1). \label{eq:phi_rec3}
\end{align} 
Note that \eqref{eq:toinvert_2}-\eqref{eq:phi_rec3} are similar to \eqref{eq:toinvert_1}-\eqref{eq:phi_rec2}, with differences due to the new definition of the extended regressor.\\

Consider again \eqref{eq:est_P}. Adding and subtracting
\begin{equation*}
\lambda_{n}\phi_{n}(T)P'\left(\rho \hat{\theta}^{g}(T-1)-\delta_{n}(T-1)\right)
\end{equation*} 
to \eqref{eq:est_P}, $\hat{\theta}_{n}^{(k+1)}$ can be computed as
\begin{align}
\nonumber \hat{\theta}_{n}^{(k+1)}(T)&=\phi_{n}(T)\left[\lambda_{n}\left(\mathcal{Y}_{n}(T-1)+P'(\rho \hat{\theta}^{g}(T-1)-\delta_{n}(T-1)\right))+\right.\\
\nonumber & \hspace{-0.7cm}\left.+X_{n}(T)y_{n}(T)-P'\left(\delta_{n}^{(k)}-\lambda_{n}\delta_{n}(T-1)\right)+P'\rho\left(\hat{\theta}^{g,(k)}-\lambda_{n}\hat{\theta}^{g}(T-1)\right)\right]=\\
&\hspace{-0.7cm} =\hat{\theta}_{n}^{RLS}(T)+\hat{\theta}_{n}^{ADMM,(k+1)}(T). \label{eq:est_dec2}
\end{align}
In particular,
\begin{align}\label{eq:est_rls3}
\nonumber \hat{\theta}_{n}^{RLS}(T)&=\phi_{n}(T)\lambda_{n}\left\{\mathcal{Y}_{n}(T-1)+\rho P' \hat{\theta}(T-1)-P'\delta_{n}(T-1)\right\}+\\
& \hspace{1cm}+\phi_{n}(T)\tilde{X}_{n}(T)y_{n}(T),
\end{align}
and
\begin{equation}\label{eq:est_admm3}
\hat{\theta}_{n}^{ADMM,(k+1)}(T)=\phi_{n}(T)P'\left(\rho \Delta_{g,\lambda_{n}}^{(k+1)}(T)- \Delta_{\lambda_{n}}^{(k+1)}\right),
\end{equation}
with
\begin{align*}
\Delta_{g,\lambda_{n}}^{k+1}(T)&=\hat{\theta}^{g,(k)}-\lambda_{n}\hat{\theta}^{g}(T-1), \\ \Delta_{\lambda_{n}}^{(k+1)}(T)&=\delta_{n}^{(k)}-\lambda_{n}\delta_{n}(T-1).
\end{align*}
Observe that, as for \eqref{eq:admm_cons:setp2} and \eqref{eq:glob_est}, \eqref{eq:est_admm3} differs from \eqref{eq:est_admm2} because of the presence of $P$.\\

Note that, accounting for the definition of $\phi_{n}(T-1)$, exploiting the equality $K_{n}(T)=\phi_{n}(T)\tilde{X}_{n}(T)$ (see Section~\ref{Sec:1} for the proof) and introducing the extended measurement vector
\begin{equation*}
\tilde{y}_{n}(T)=\begin{bmatrix}y_{n}(T)' & O_{1\times n_{g}}\end{bmatrix}',
\end{equation*}
the formula to update $\hat{\theta}_{n}^{RLS}$ in \eqref{eq:est_admm3} can be further simplified as
\begin{align}
\nonumber&\hat{\theta}_{n}^{RLS}(T)=\phi_{n}(T-1)\left\{ \left(\mathcal{Y}_{n}(T-1)+P'(\rho\hat{\theta}^{g}(T-1)-\delta_{n}(T-1))\right)\right\}+\\
\nonumber &\hspace{0.5cm}-K_{n}(T)(\tilde{X}_{n}(T)')\phi_{n}(T-1)\left\{\left(\mathcal{Y}_{n}(T-1)+P'(\rho\hat{\theta}^{g}(T-1)-\delta_{n}(T-1))\right)\right\}+\\
\nonumber &\hspace{0.5cm}+\phi_{n}(T)X_{n}(T)y_{n}(T)=\\ 
\nonumber &\hspace{0.5cm}=\hat{\theta}_{n}(T-1)-K_{n}(T)(\tilde{X}_{n}(T))'\hat{\theta}_{n}(T-1)+\phi_{n}(T)\tilde{X}_{n}(T)\tilde{y}_{n}(T)=\\
&\hspace{0.5cm}=\hat{\theta}_{n}(T-1)+K_{n}(T)(\tilde{y}_{n}(T)-(\tilde{X}_{n}(T))'\hat{\theta}_{n}(T-1)). \label{eq:rls_par}
\end{align}
As the method tailored to attain full consensus (see Section~\ref{Sec:1}), note that both $\hat{\theta}^{g}$ and $\delta_{n}$ should be updated on the \textquotedblleft cloud\textquotedblright. As a consequence, also $\hat{\theta}_{n}^{ADMM}$ should be updated on the \textquotedblleft cloud\textquotedblright, due to its dependence on both $\hat{\theta}^{g}$ and $\delta_{n}$. On the other hand, $\hat{\theta}_{n}^{RLS}$ can be updated by the local processors. As for the case considered in Section~\ref{Sec:1}, note that \eqref{eq:rls_par} is independent from $k$ and, consequently, the synchronization between the local clock and the one on the \textquotedblleft cloud\textquotedblright is not required.\\

The approach is outlined in Algorithm~\ref{algo5} and the transmissions characterizing each iteration is still the one reported in the scheme in \figurename{~\ref{Fig:ADMMscheme}}. As a consequence, the observations made in Section~\ref{Sec:1} with respect to the information exchange between the nodes and the \textquotedblleft cloud\textquotedblright \ hold also in this case.

\begin{algorithm}[!tb]
	\caption{ADMM-RLS for partial consensus (N2C2N)}
	\label{algo5}
	~~\textbf{Input}: Sequence of observations $\{X_{n}(t),y_{n}(t)\}_{t=1}^T$, initial matrices $\phi_{n}(0) \in \mathbb{R}^{n_{\theta} \times n_{\theta}}$, initial local estimates $\hat{\theta}_{n}(0)$, initial dual variables $\delta_{n,\mathrm{o}}$, forgetting factors $\lambda_{n}$, $n=1,\ldots,N$, initial global estimate $\hat{\theta}_{\mathrm{o}}^{g}$, parameter $\rho \in \mathbb{R}^{+}$.
	\vspace*{.1cm}\hrule\vspace*{.1cm}
	\begin{enumerate}[label=\arabic*., ref=\theenumi{}]  
		\item \textbf{for} $t=1,\ldots,T$ \textbf{do}
		\begin{itemize}
			\item[] \hspace{-0.5cm} \textbf{\underline{Local}}
			\begin{enumerate}[label=\theenumi{}.\arabic*., ref=\theenumi{}.\theenumii{}]
				\item \textbf{for} $n=1,\ldots,N$ \textbf{do}
				\begin{enumerate}[label=\theenumii{}.\arabic*., ref=\theenumi{}.\theenumii{}.\theenumiii{}]
					\item \textbf{compute} $\tilde{X}_{n}(t)$ with \eqref{eq:tildeX_2};
					\item \textbf{compute} $K_{n}(t)$ and $\phi_{n}(t)$ with \eqref{eq:gain_3}~-~\eqref{eq:phi_rec3};
					\item \textbf{compute} $\hat{\theta}_{n}^{RLS}(t)$ with \eqref{eq:rls_par};
				\end{enumerate}
				\item \textbf{end for};
			\end{enumerate}
			\item[] \hspace{-0.5cm} \textbf{\underline{Global}}
			\begin{enumerate}[label=\theenumi{}.\arabic*., ref=\theenumi{}.\theenumii{}]
				\item \textbf{do} 
				\begin{enumerate}[label=\theenumii{}.\arabic*., ref=\theenumi{}.\theenumii{}.\theenumiii{}]
					\item \textbf{compute} $\hat{\theta}_{n}^{ADMM,(k+1)}(t)$ with \eqref{eq:est_admm3}, $n=1,\ldots,N$;
					\item \textbf{compute} $\hat{\theta}_{n}^{(k+1)}(t)$ with \eqref{eq:est_dec2}, $n=1,\ldots,N$;
					\item \textbf{compute} $\hat{\theta}^{g,(k+1)}$ with \eqref{eq:glob_est};
					\item \textbf{compute} $\delta_{n}^{(k+1)}$ with \eqref{step:dual_est}, $n=1,\ldots,N$;
				\end{enumerate}
				\item \textbf{until} a stopping criteria is satisfied (e.g. maximum number of iterations attained);
			\end{enumerate}
		\end{itemize}
		\item \textbf{end}.
	\end{enumerate}
	\vspace*{.1cm}\hrule\vspace*{.1cm}
	~~\textbf{Output}: Estimated global parameters $\{\hat{\theta}^{g}(t)\}_{t=1}^{T}$, estimated local parameters $\{\hat{\theta}_{n}(t)\}_{t=1}^{T}$, $n=1,\ldots,N$.
\end{algorithm}
\subsection{Example 3}
Assume to collect data for $T=1000$ from a set of $N=100$ dynamical systems modelled as
\begin{equation}\label{eq:syst3}
y_{n}(t)=\theta_{1}^{g}y_{n}(t-1)+\theta_{n,2}y_{n}(t-2)+\theta_{2}^{g}u_{n}(t-1)+e_{n}(t),
\end{equation}
where $\theta^{g}=\smallmat{0.2 & 0.8}'$ and $\theta_{n,2}$ is sampled from a normal distribution $\mathcal{N}(0.4,0.0025)$, so that it is different for the $N$ systems. The  white noise sequence $e_{n}\sim \mathcal{N}(0,R_{n})$, where, for the \textquotedblleft informative\textquoteright \ systems, $R_{n} \in [1 \ 20]$ yields $SNR \in [3.1,14.6]$~dB (see~\eqref{eq:snr}).\\
Initializing $\phi_{n}$ as $\phi_{n}(0)=0.1I_{n_{\theta}}$, while $\hat{\theta}_{n}(0)$ and $\hat{\theta}_{\mathrm{o}}^{g}$ are sampled from the distributions $\mathcal{N}(\hat{\theta}^{g},2I_{n_{\theta}})$ and $\mathcal{N}(\hat{\theta}^{g},I_{n_{\theta}})$, respectively, $\{\lambda_{n}=\Lambda\}_{n=1}^{N}$, with $\Lambda=1$, and $\rho=0.1$, the performance of the proposed approach are evaluated. \figurename{~\ref{Fig:globParams3}} shows $\hat{\theta}^{g}$ obtained with ADMM-RLS, along with the estimation error. Observe that the estimates tends to converge to the actual value of the global parameters.
\begin{figure}[!tb]
	\centerline{
		\begin{tabular}[t]{cc}
			\subfigure[$\theta_{1}^{g}$ vs $\hat{\theta}_{1}^{g}$]{\includegraphics[scale=0.7]{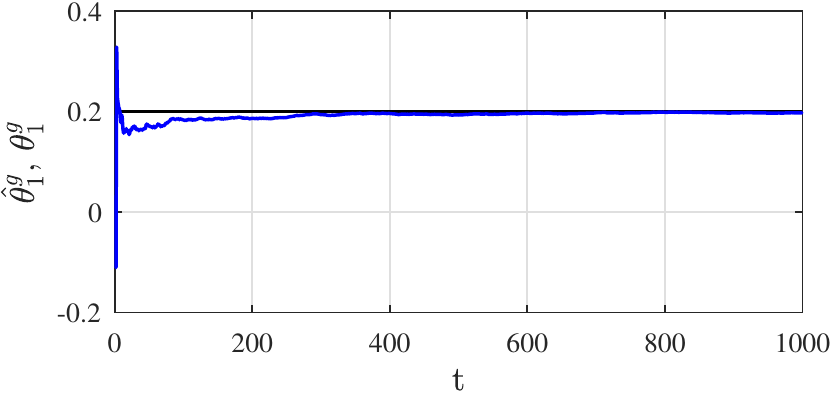}}
			\subfigure[$|\hat{\theta}_{1}^{g}-\theta_{1}^{g}|$]{\includegraphics[scale=0.7]{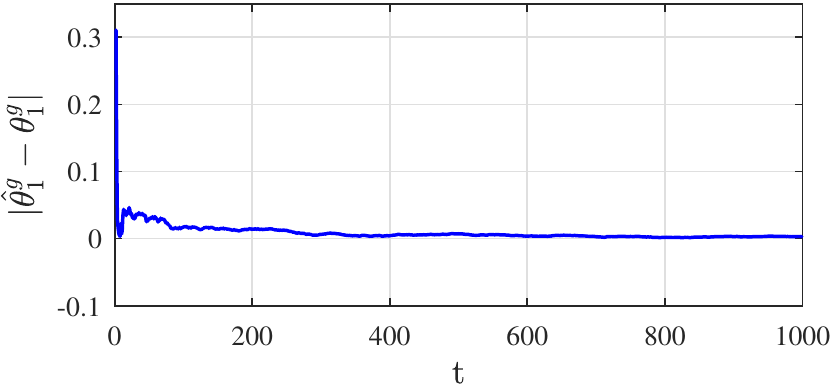}}\\
			\subfigure[$\theta_{2}^{g}$ vs $\hat{\theta}_{2}^{g}$]{\includegraphics[scale=0.7]{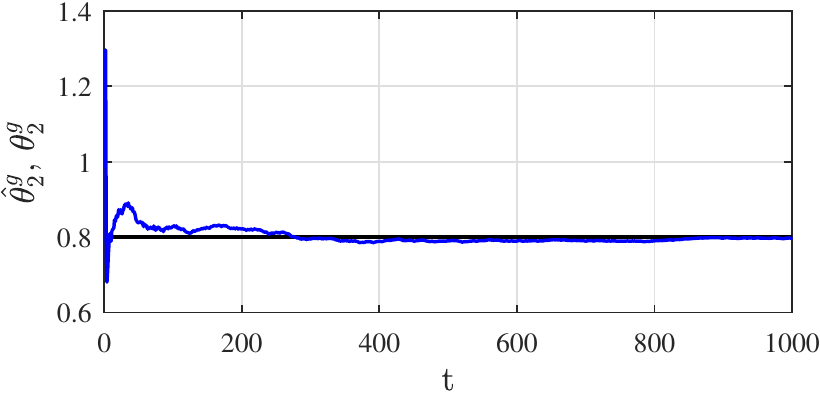}}
			\subfigure[$|\hat{\theta}_{2}^{g}-\theta_{2}^{g}|$]{\includegraphics[scale=0.7]{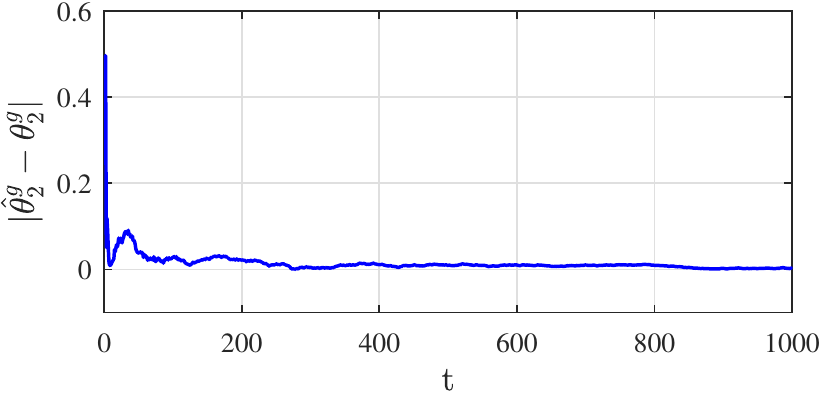}}\\
		\end{tabular}
	}
	\caption{Example 3. True vs estimated global parameters. Black : true, blue : ADMM-RLS.}
	\label{Fig:globParams3}
\end{figure}
To further assess the performances of ADMM-RLS, $\theta_{n}$, $\hat{\theta}_{n}$ and $\hat{\theta}_{n}^{RLS}$ obtained for the $5$th system, i.e. $n=5$, are compared in \figurename{~\ref{Fig:localParams3}}. It can thus be seen that the difference between $\hat{\theta}_{n}^{RLS}$ and $\hat{\theta}_{n}$ is mainly noticeable at the beginning of the estimation horizon, but then $\hat{\theta}_{n}^{RLS}$ and $\hat{\theta}_{n}$ are barely distinguishable. Note that $\mbox{SNR}_{5}=8.9$~dB. 
\begin{figure}[!tb]
		\begin{tabular}[t]{cc}
			\subfigure[$\theta_{5,1}$ vs $\hat{\theta}_{5,1}$ and $\hat{\theta}_{5,1}^{RLS}$]{\includegraphics[scale=0.7]{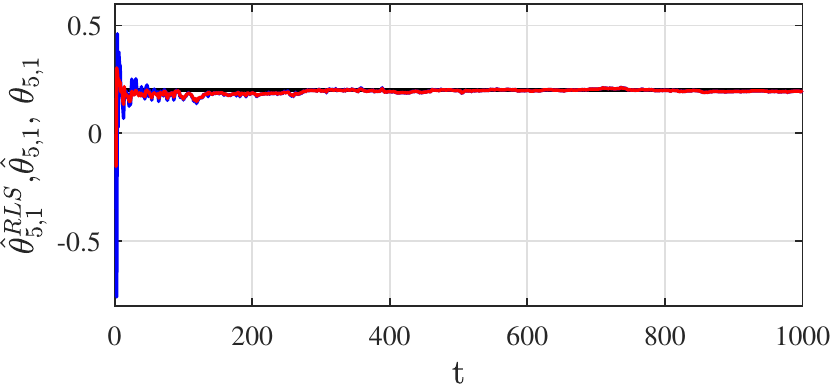}}
			\subfigure[$\theta_{5,2}$ vs $\hat{\theta}_{5,2}$ and $\hat{\theta}_{5,2}^{RLS}$]{\includegraphics[scale=0.7]{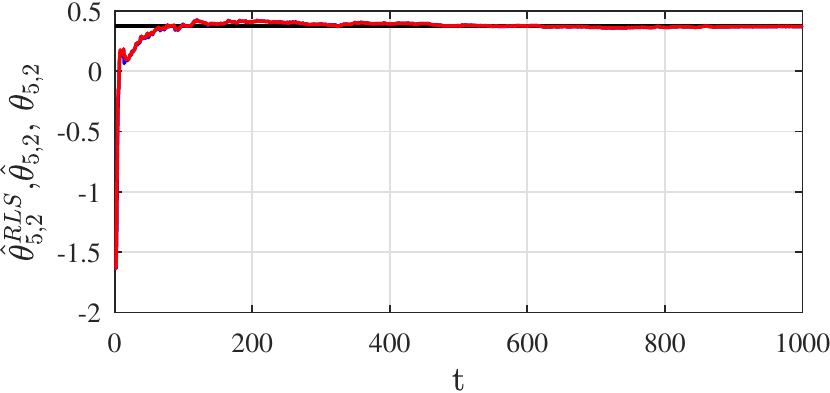}}\\
			\multicolumn{1}{c}{\subfigure[$\theta_{5,3}$ vs $\hat{\theta}_{5,3}$ and $\hat{\theta}_{5,3}^{RLS}$]{\includegraphics[scale=0.7]{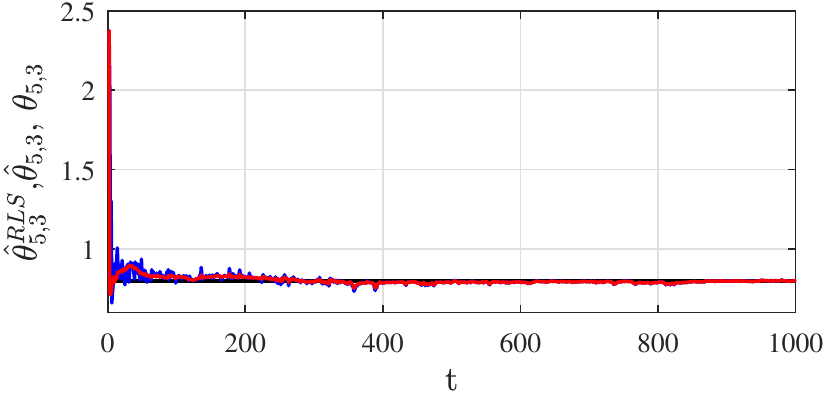}}}
		\end{tabular}
	\caption{Example 3. Local parameter $\theta_{n,2}$, $n=5$. Black : true,
		blue : $\hat{\theta}_{5}$, red: $\hat{\theta}_{5}^{RLS}$.}
	\label{Fig:localParams3}
\end{figure}
\subsubsection{Non-informative agents}
Suppose that among the $N=100$ systems described by the model in \eqref{eq:syst3}, $N_{ni}=20$ randomly chosen agents are non-informative, i.e. their input sequences $u_{n}$ are null and $R_{n}=10^{-8}$.\\
As it can be observed from the estimates reported in \figurename{~\ref{Fig:pramsGreedyADMM3}}, $\{\hat{\theta}_{i}^{g}\}_{i=1}^{2}$ converge to the actual values of the global parameters even if $20$\% of the systems provide non-informative data.\\
\begin{figure}[!tb]
	\centerline{
		\begin{tabular}[!tb]{cc}
			\subfigure[$\theta_{1}^{g}$ vs $\hat{\theta}_{1}^{g}$]{\includegraphics[scale=0.7]{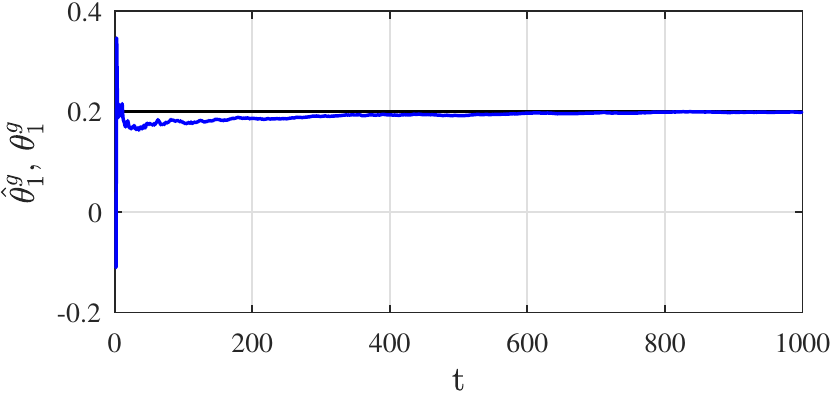}}
			\subfigure[$\theta_{2}^{g}$ vs $\hat{\theta}_{2}^{g}$]{\includegraphics[scale=0.7]{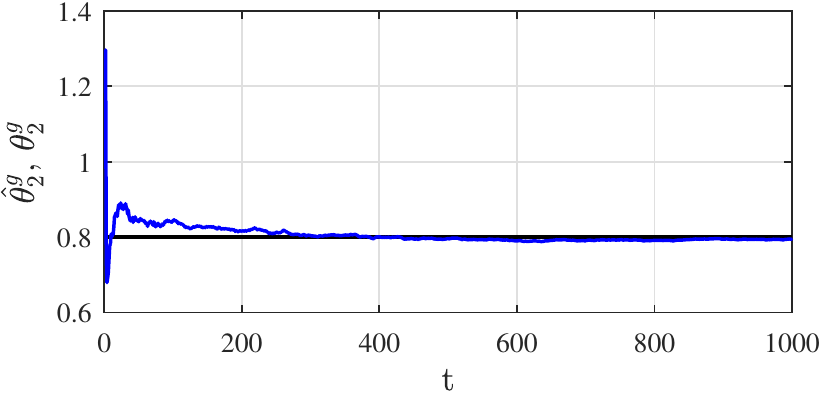}}\\
		\end{tabular}
	}
	\caption{Example 3. True vs estimated global parameters. Black : true, blue : ADMM-RLS.}
	\label{Fig:pramsGreedyADMM3}
\end{figure}
The local estimates $\hat{\theta}_{n,2}$ for the $8$th and $65$th system ($SNR_{65}\approx6$~dB) are reported in \figurename{~\ref{Fig:prams8_65_notexc}. As, the $8$th system is among the ones with a non exciting input, $\hat{\theta}_{8,2}=\hat{\theta}_{8,2}(0)$ over the estimation horizon. Instead, $\hat{\theta}_{65,2}$ tends to converge to the actual value of $\theta_{65,2}$. Even if the purely local parameter is not retrieved from the data, using the proposed collaborative approach $\theta_{8,1}$ and $\theta_{8,3}$ are accurately estimated (see~\figurename{~\ref{Fig:prams8_notexc}}). We can thus conclude that the proposed estimation method \textquotedblleft forces\textquotedblright \ the estimates of the global components of $\theta_{n}$ to follow $\hat{\theta}^{g}$, which is estimated automatically discarding the contributions from the systems that lacked excitation.
\begin{figure}[!tb]
	\centerline{
		\begin{tabular}[!tb]{cc}
			\subfigure[$\theta_{8,2}$ vs $\hat{\theta}_{8,2}$ \label{Fig:sub_loc8}]{\includegraphics[scale=0.7]{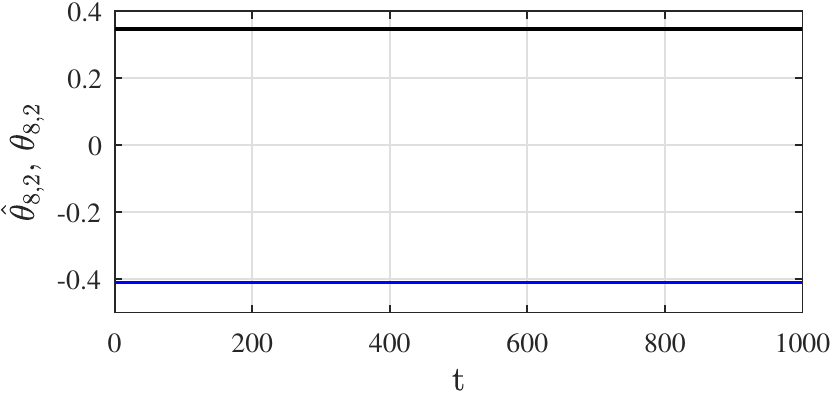}} 
			\subfigure[$\theta_{65,2}$ vs $\hat{\theta}_{65,2}$\label{Fig:sub_loc65}]{\includegraphics[scale=0.7]{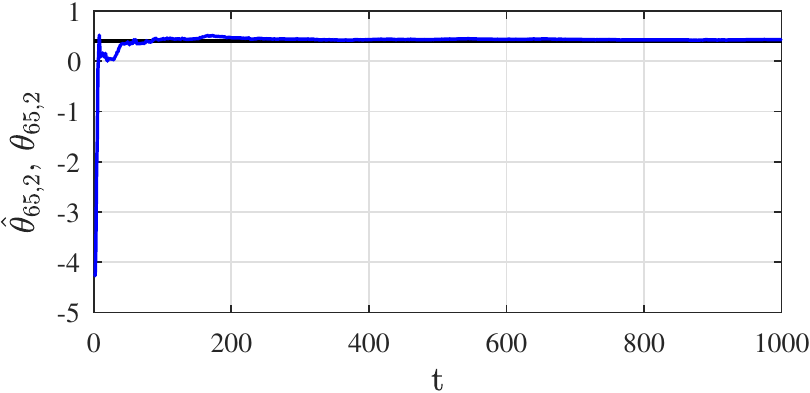}}
		\end{tabular}
	}
	\caption{Example 3. Local parameters $\theta_{n,2}$, $n=8,65$. Black : true, blue : ADMM-RLS.}
	\label{Fig:prams8_65_notexc}
\end{figure}
\begin{figure}[!tb]
	\centerline{
		\begin{tabular}[!tb]{cc}
			\subfigure[$\theta_{8,1}$ vs $\hat{\theta}_{8,1}$ and $\hat{\theta}_{8,1}^{RLS}$ ]{\includegraphics[scale=0.7]{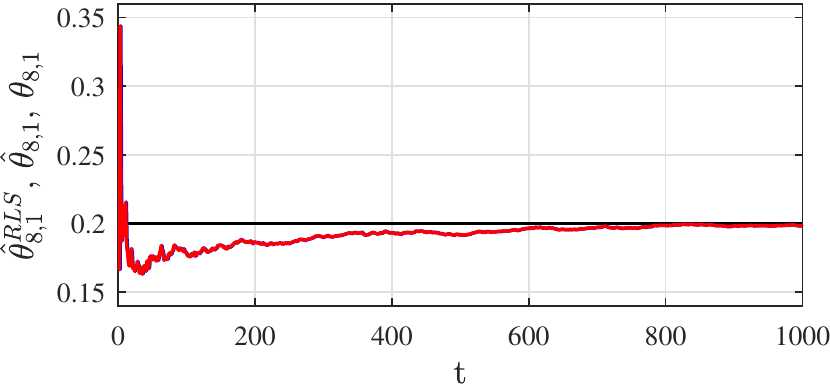}} &
			\subfigure[$\theta_{8,3}$ vs $\hat{\theta}_{8,3}$ and $\hat{\theta}_{8,3}^{RLS}$]{\includegraphics[scale=0.7]{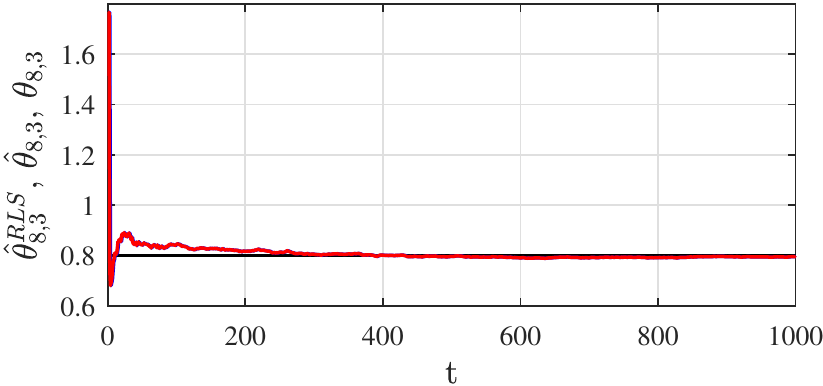}}\\
		\end{tabular}
	}
	\caption{Example 3. Local parameters $\theta_{8,i}$, $i=1,3$. Black : true, blue : $\hat{\theta}_{8,i}^{RLS}$, red: $\hat{\theta}_{8,i}$.}
	\label{Fig:prams8_notexc}
\end{figure}
\section{Constrained Collaborative estimation for partial consensus}\label{Sec:3}
Suppose that the value of the local parameter $\theta_{n}$ is constrained to a set $\mathcal{C}_{n}$ and that this hypothesis holds for all the agents $n \in \{1,\ldots,N\}$. With the objective of reaching partial consensus among the agents, the problem to be solved can thus be formulated as
\begin{equation}\label{eq:prob_const}
\begin{aligned}
&\mbox{minimize } && \sum_{n=1}^{N} f_{n}(\theta_{n})\\
&\mbox{s.t. }&&P\theta_{n}=\theta, \ \ n=1,\ldots,N,\\
&&& \theta_{n} \in \mathcal{C}_{n}, \ \ n=1,\ldots,N.
\end{aligned}
\end{equation}
Observe that \eqref{eq:prob_const} corresponds to \eqref{eq:problem} if the nonlinear consensus constraint is replaced with \eqref{eq:partial_constr}.\\
To use ADMM to solve \eqref{eq:prob_const}, the problem has to be modified as
\begin{equation}\label{eq:prob_const2}
\begin{aligned}
&\mbox{minimize } &&\sum_{n=1}^{N} \left\{f_{n}(\theta_{n}) + g_{n}(z_{n})\right\}\\
&  \mbox{s.t. } && P\theta_{n}=\theta^{g} \ \ n=1,\ldots,N\\
&&& \theta_{n}=z_{n}, \ \ n=1,\ldots,N
\end{aligned}
\end{equation}
where $\{g_{n}\}_{n=1}^{N}$ are the indicator functions of the sets $\{\mathcal{C}_{n}\}_{n=1}^{N}$ (defined as in \eqref{eq:ind_func}) and $\{z_{n} \in \mathbb{R}^{n_{\theta}}\}_{n=1}^{N}$ are auxiliary variables. Observe that \eqref{eq:prob_const2} can be solved with ADMM. Given the augmented Lagrangian associated with \eqref{eq:prob_const2}, i.e.
\begin{align}\label{eq:lag_const}
\nonumber \mathcal{L}&=\sum_{n=1}^{N}\{f_{n}(\theta_{n})+g_{n}(z_{n})+\delta_{n,1}'(\theta_{n}-z_{n})+\delta_{n,2}'(P\theta_{n}-\theta^{g})+\\
&\hspace{1cm}+\frac{\rho_{1}}{2}\|\theta_{n}-z_{n}\|_{2}^{2}+\frac{\rho_{2}}{2}\|P\theta_{n}-\theta^{g}\|_{2}^{2}\},
\end{align}
the iterations that have to be performed to solve the addressed problem with ADMM are
\begin{align}
&\hat{\theta}_{n}^{(k+1)}(T)=\underset{\theta_{n}}{\argmin} \ \mathcal{L}(\theta_{n},\hat{\theta}^{g,(k)},z_{n}^{(k)},\delta_{n}^{(k)}),
\label{step:loc_est2}\\
&z_{n}^{(k+1)}=\underset{z_{n}}{\argmin} \ \mathcal{L}(\hat{\theta}_{n,(k+1)}(T),\hat{\theta}^{g,(k)},z_{n},\delta_{n}^{(k)}),
\label{step:aux_est2}\\
&\hat{\theta}^{g,(k+1)}=\underset{\theta^{g}}{\argmin} \ \mathcal{L}(\{\hat{\theta}_{n}^{(k+1)}\}_{n=1}^{N},\theta^{g},\{z_{n}^{(k+1)},\delta_{n}^{(k)}\}_{n=1}^{N}), \label{step:glob_est2}\\
&\delta_{n,1}^{(k+1)}=\delta_{n,1}^{(k)}+\rho_{1}(\hat{\theta}_{n}^{(k+1)}(T)-z^{(k+1)}),\label{eq:dual1_update}\\
&\delta_{n,2}^{(k+1)}=\delta_{n,2}^{(k)}+\rho_{2}(P\hat{\theta}_{n}^{(k+1)}(T)-\hat{\theta}^{g,(k+1)}).\label{eq:dual2_update}
\end{align}
Note that two sets of Lagrangian multipliers, $\{\delta_{n,1}\}_{n=1}^{N}$ and $\{\delta_{n,2}\}_{n=1}^{N}$, have been introduced. While $\delta_{n,1} \in \mathbb{R}^{n_g}$ is associated with the partial consensus constraint, $\delta_{n,2} \in \mathbb{R}^{n_{\theta}}$ is related to the constraint $\theta_{n} \in \mathcal{C}_{n}$, $n=1,\ldots,N$.\\

Solving \eqref{step:aux_est2}-\eqref{step:glob_est2}, the resulting updates for the auxiliary variables and the global estimates are
\begin{align}
z_{n}^{(k+1)}=&\mathcal{P}_{\mathcal{C}_{n}}\left(\hat{\theta}_{n}^{(k+1)}(T)+\frac{1}{\rho_{1}}\delta_{n,1}^{(k)}\right), \ \ n=1,\ldots,N, \label{eq:z_update}\\
\hat{\theta}^{g,(k+1)}&=\frac{1}{N}\sum_{n=1}^{N}\left(P\hat{\theta}_{n}^{(k+1)}(T)+\frac{1}{\rho_{2}}\delta_{n,2}^{(k)} \right). \label{eq:thetag_update}
\end{align}
Observe that $z$-update is performed projecting onto the set $\mathcal{C}_{n}$ a combination of the updated local estimate and $\delta_{n,1}^{(k)}$, while $\hat{\theta}^{g,(k+1)}$ is computed as in Section~\ref{Sec:2}, with $\delta_{n}$ replaced by $\delta_{n,2}$.\\

Consider the close form solution of \eqref{step:loc_est2}, which is given by
\begin{align}
\hat{\theta}_{n}^{(k+1)}(T)&=\phi_{n}(T)\left\{\mathcal{Y}_{n}(T)-\delta_{n,1}^{(k)}-P'\delta_{n,2}^{(k)}+\rho_{1}z_{n}^{(k)}+\rho_{2}P'\hat{\theta}^{g,(k)} \right\}, \label{eq:explicit1}\\
\mathcal{Y}_{n}(t)&=\sum_{\tau=1}^{t}\lambda_{n}^{t-\tau} X_{n}(\tau)y_{n}(\tau),\\
\phi_{n}(t)&=\left(\left[\sum_{\tau=1}^{t}\lambda_{n}^{t-\tau} X_{n}(\tau)X_{n}(\tau)'\right]+\rho_{1}I_{n_{\theta}}+\rho_{2}P'P\right)^{-1}. \label{eq:phi_3}
\end{align} 
Aiming at finding recursive formulas to update the estimates of the local parameters, we introduce the $n$th local estimate obtained at $T-1$, i.e.
\begin{align}\label{eq:prev_1}
\nonumber \hat{\theta}_{n}(T-1)&=\phi_{n}(T-1)\left\{\mathcal{Y}_{n}(T-1)-\delta_{n,1}(T-1)-P'\delta_{n,2}(T-1)+\right.\\
&\hspace{1cm}\left.+\rho_{1}z_{n}(T-1)+\rho_{2}P'\hat{\theta}^{g}(T-1) \right\}
\end{align}
with $\delta_{n,1}(T-1)$, $\delta_{n,2}(T-1)$, $z_{n}(T-1)$ and $\hat{\theta}^{g}(T-1)$ being the Lagrange multipliers and the global estimate obtained at $T-1$, respectively.\\

To obtain recursive formulas to compute $\hat{\theta}_{n}^{(k+1)}$, we start proving that $\phi_{n}(T)$ can be computed as a function of $\phi_{n}(T-1)$. in particular, introducing
\begin{equation*}
\mathcal{X}_{n}(t)=\sum_{\tau=1}^{t}\lambda_{n}^{t-\tau} X_{n}(\tau)(X_{n}(\tau))',
\end{equation*}
note that
\begin{align*}
&\phi_{n}(T)^{-1}=\mathcal{X}_{n}(T)+\rho_{1}I_{n_{\theta}}+\rho_{2}P'P=\\
&=\lambda_{n}\mathcal{X}_{n}(T-1)+X_{n}(T)X_{n}(T)'+\rho_{1}I_{n_{\theta}}+\rho_{2}P'P=\\
&=\lambda_{n}\left[\mathcal{X}_{n}(T-1)+\rho_{1}I_{n_{\theta}}+\rho_{2}P'P\right]+X_{n}(T)X_{n}(T)'+(1-\lambda_{n})\rho_{1}+(1-\lambda_{n})\rho_{2}P'P=\\
&=\lambda_{n}\phi_{n}(T-1)^{-1}+X_{n}(T)X_{n}(T)'+(1-\lambda_{n})\rho_{1}+(1-\lambda_{n})\rho_{2}P'P.
\end{align*}
Defining the extended regressor as
\begin{equation}\label{eq:tildeX_3}
\tilde{X}_{n}(T)=\begin{bmatrix}
X_{n}(T) & \sqrt{(1-\lambda_{n})\rho_{1}}I_{n_{\theta}} & \sqrt{(1-\lambda_{n})\rho_{2}}P' 
\end{bmatrix} \in \mathbb{R}^{n_{\theta}\times(n_{y}+n_{\theta}+n_{g})},
\end{equation}
and applying the matrix inversion lemma, it can be easily proven that $\phi_{n}(T)$ can then be computed as:
\begin{align}
\mathcal{R}_{n}(T)&=\lambda_{n}I_{(n_{y}+n_{\theta}+n_{g})}+\tilde{X}_{n}(T)'\phi_{n}(T)\tilde{X}_{n}(T),\\
K_{n}(T)&=\phi_{n}(T-1)\tilde{X}_{n}(T)(\mathcal{R}_{n}(T))^{-1},\label{eq:gain_4}\\
\phi_{n}(T)&=\lambda_{n}^{-1}(I_{n_{\theta}}-K_{n}(T)\tilde{X}_{n}(T)')\phi_{n}(T-1).\label{eq:phi_rec4}
\end{align}
The same observations relative to the update of $\phi_{n}$ made in Section~\ref{Sec:2} holds also in the considered case.\\

Consider \eqref{eq:explicit1}. Adding and subtracting 
 \begin{equation*}
 \lambda_{n}\left[-\delta_{n,1}(T-1)-P'\delta_{n,2}(T-1)+\rho_{1}z_{n}(T-1)+\rho_{2}P'\hat{\theta}^{g}(T-1)\right]
 \end{equation*}
 to \eqref{eq:explicit1} and considering the definition of $\phi_{n}(T-1)$ (see~\eqref{eq:phi_3}), the formula to update $\hat{\theta}_{n}$ can be further simplified as
 \begin{align} \nonumber &\hat{\theta}_{n}^{(k+1)}(T)=\phi_{n}(T)\{\lambda_{n}\left(\mathcal{Y}_{n}(T-1)-\delta_{n,1}(T-1)-P'\delta_{n,2}(T-1)+\right.\\
\nonumber  &\hspace{0.2cm}\left.+\rho_{1}z_{n}(T-1)+\rho_{2}P'\hat{\theta}^{g}(T-1)\right) +X_{n}(T)y_{n}(T)+\rho_{1}(z_{n}^{(k)}-\lambda_{n}z_{n}(T-1))+\\
\nonumber&\hspace{0.2cm}+\rho_{2}P'(\hat{\theta}^{g,(k)}-\lambda_{n}\hat{\theta}^{g}(T-1))-(\delta_{n,1}^{(k)}-\lambda_{n}\delta_{n,1}(T-1))+\\
\nonumber&\hspace{0.2cm}-P'(\delta_{n,2}^{(k)}-\lambda_{n}\delta_{n,2}(T-1))\}=\\
\nonumber &\hspace{0cm}=\hat{\theta}_{n}(T-1)-K_{n}(T)\tilde{X}_{n}(T)\hat{\theta}_{n}(T-1)+\phi_{n}(T)\{X_{n}(T)y_{n}(T)+\\
\nonumber &\hspace{0.2cm} +\rho_{1}(z_{n}^{(k)}-\lambda_{n}z_{n}(T-1))+\rho_{2}P'(\hat{\theta}^{g,(k)}-\lambda_{n}\hat{\theta}^{g}(T-1))\}+\\
\nonumber & \hspace{0.2cm} -(\delta_{n,1}^{(k)}-\lambda_{n}\delta_{n,1}(T-1))-P'(\delta_{n,2}^{(k)}-\lambda_{n}\delta_{n,2}(T-1))\}=\\
&= \hat{\theta}_{n}^{RLS}(T)+\hat{\theta}_{n}^{ADMM,(k+1)}(T). \label{eq:est_dec3}
 \end{align}
In particular,
\begin{align}\label{eq:est_rls4} \nonumber\hat{\theta}_{n}^{RLS}&=\phi_{n}(T)\lambda_{n}\left(\mathcal{Y}_{n}(T-1) -\delta_{n,1}(T-1)-P'\delta_{n,2}(T-1)+\rho_{1}z_{n}(T-1)+\right.\\
&\hspace{1cm}\left.+\rho_{2}P'\hat{\theta}^{g}(T-1)\right)+\phi_{n}(T)X_{n}(T)y_{n}(T),
\end{align}
while
\begin{equation}\label{eq:est_admm4}
\hat{\theta}_{n}^{ADMM,(k+1)}(T)=\phi_{n}(T)\left[\rho_{1}\Delta_{z,\lambda_{n}}^{(k+1)}(T)+\rho_{2}P' \Delta_{g,\lambda_{n}}^{(k+1)}(T)- \Delta_{1,\lambda_{n}}^{(k+1)}-P'\Delta_{2,\lambda_{n}}^{(k+1)}\right].
\end{equation}
with
\begin{align*} &\Delta_{z,\lambda_{n}}^{(k+1)}(T)=z_{n}^{(k)}-\lambda_{n}z_{n}(T-1),\\  &\Delta_{g,\lambda_{n}}^{(k+1)}(T)=\hat{\theta}^{g,(k)}-\lambda_{n}\hat{\theta}^{g}(T-1),\\ &\Delta_{1,\lambda_{n}}^{(k+1)}=\delta_{n,1}^{(k)}-\lambda_{n}\delta_{n,1}(T-1),\\ 
&\Delta_{2,\lambda_{n}}^{(k+1)}=\delta_{n,2}^{(k)}-\lambda_{n}\delta_{n,2}(T-1).
\end{align*}
Note that \eqref{eq:est_admm4} differs from \eqref{eq:est_admm3} because of the introduction of the additional terms $\Delta_{z,\lambda_{n}}$ and $\Delta_{1,\lambda_{n}}$.\\
Similarly to what is presented in Section~\ref{Sec:2}, thanks to \eqref{eq:phi_rec4} the formula to update $\hat{\theta}_{n}^{RLS}$ can be further reduced as
\begin{align*} 
\hat{\theta}_{n}^{RLS}&=\hat{\theta}_{n}(T-1)-K_{n}(T)(\tilde{X}_{n}(T))'\hat{\theta}_{n}(T-1)+\phi_{n}(T)X_{n}(T)y_{n}(T)=\\
&=\hat{\theta}_{n}(T-1)-K_{n}(T)(\tilde{X}_{n}(T))'\hat{\theta}_{n}(T-1)+\phi_{n}(T)\tilde{X}_{n}(T)\tilde{y}_{n}(T),
\end{align*}
with the extended measurement vector $\tilde{y}_{n}(T)$ is defined as
\begin{equation*}
\tilde{y}_{n}(T)=\begin{bmatrix}y_{n}(T)' & O_{1\times n_{\theta}} & O_{1\times n_{n_{g}}}\end{bmatrix}'.
\end{equation*}
Exploiting the equality $K_{n}(T)=\phi_{n}(T)\tilde{X}_{n}(T)$ (the proof can be found in \eqref{Sec:1}), it can thus be proven that
\begin{equation}\label{eq:rls_parC} 
\hat{\theta}_{n}^{RLS}=hat{\theta}_{n}(T-1)+K_{n}(T)(\tilde{y}_{n}(T)-(\tilde{X}_{n}(T))'\hat{\theta}_{n}(T-1)).
\end{equation}
It is worth remarking that $\hat{\theta}_{n}^{RLS}$ can be updated ($i$) locally, ($ii$) recursively and ($iii$) once per step $t$.
\begin{algorithm}[!b]
	\caption{ADMM-RLS algorithm for constrained consensus}
	\label{algo6}
	~~\textbf{Input}: Sequence of observations $\{X_{n}(t),y_{n}(t)\}_{t=1}^T$, initial matrices $\phi_{n}(0) \in \mathbb{R}^{n_{\theta} \times n_{\theta}}$, initial local estimates $\hat{\theta}_{n}(0)$, initial dual variables $\delta_{n,1}^{\mathrm{o}}$ and $\delta_{n,2}^{\mathrm{o}}$, initial auxiliary variables $\hat{z}_{n,\mathrm{o}}$, forgetting factors $\lambda_{n}$, $n=1,\ldots,N$, initial global estimate $\hat{\theta}_{\mathrm{o}}^{g}$, parameters $\rho_{1}, \rho_{2} \in \mathbb{R}^{+}$.
	\vspace*{.1cm}\hrule\vspace*{.1cm}
	\begin{enumerate}[label=\arabic*., ref=\theenumi{}]  
		\item \textbf{for} $t=1,\ldots,T$ \textbf{do}
		\begin{itemize}
			\item[] \hspace{-0.5cm} \textbf{\underline{Local}}
			\begin{enumerate}[label=\theenumi{}.\arabic*., ref=\theenumi{}.\theenumii{}]
				\item \textbf{for} $n=1,\ldots,N$ \textbf{do}
				\begin{enumerate}[label=\theenumii{}.\arabic*., ref=\theenumi{}.\theenumii{}.\theenumiii{}]
					\item \textbf{compute} $\tilde{X}_{n}(t)$ with \eqref{eq:tildeX_3};					
					\item \textbf{compute} $K_{n}(t)$ and $\phi_{n}(t)$ with \eqref{eq:gain_4}~-~\eqref{eq:phi_rec4};
					\item \textbf{compute} $\hat{\theta}_{n}^{RLS}(t)$ with \eqref{eq:rls_parC};
				\end{enumerate}
				\item \textbf{end for};
			\end{enumerate}
			\item[] \hspace{-0.5cm} \textbf{\underline{Global}}
			\begin{enumerate}[label=\theenumi{}.\arabic*., ref=\theenumi{}.\theenumii{}]
				\item \textbf{do} 
				\begin{enumerate}[label=\theenumii{}.\arabic*., ref=\theenumi{}.\theenumii{}.\theenumiii{}]
					\item \textbf{compute} $\hat{\theta}_{n}^{ADMM,(k+1)}(t)$ with \eqref{eq:est_admm4}, $n=1,\ldots,N$;
					\item \textbf{compute} $\hat{\theta}_{n}^{(k+1)}(t)$ with \eqref{eq:est_dec3}, $n=1,\ldots,N$;
					\item \textbf{compute} $z_{n}^{(k+1)}(t)$ with \eqref{eq:z_update}, $n=1,\ldots,N$;
					\item \textbf{compute} $\hat{\theta}^{g,(k+1)}$ with \eqref{eq:thetag_update};
					\item \textbf{compute} $\delta_{n,1}^{(k+1)}$ with \eqref{eq:dual1_update}, $n=1,\ldots,N$;
					\item \textbf{compute} $\delta_{n,2}^{(k+1)}$ with \eqref{eq:dual2_update}, $n=1,\ldots,N$;
				\end{enumerate}
				\item \textbf{until} a stopping criteria is satisfied (e.g. maximum number of iterations attained);
			\end{enumerate}
		\end{itemize}
		\item \textbf{end}.
	\end{enumerate}
	\vspace*{.1cm}\hrule\vspace*{.1cm}
	~~\textbf{Output}: Estimated global parameters $\{\hat{\theta}^{g}(t)\}_{t=1}^{T}$, estimated local parameters $\{\hat{\theta}_{n}(t)\}_{t=1}^{T}$, $n=1,\ldots,N$.
\end{algorithm}
\begin{remark}
	The proposed method, summarized in Algorithm~\ref{algo6} and in \figurename{~\ref{Fig:ADMMscheme}}, requires the agents to transmit $\{\hat{\theta}_{n}^{RLS},\phi_{n}\}$ to the \textquotedblleft cloud\textquotedblright, while the \textquotedblleft cloud\textquotedblright \ has to communicate $\hat{\theta}_{n}$ to each node once it has been computed. As a consequence, a N2C2N transmission scheme is required. \hfill $\blacksquare$
\end{remark}

\subsection{Example 4}
Suppose that the data are gathered from $N=100$ systems, described by \eqref{eq:syst3} and  collected over an estimation horizon $T=5000$. Moreover, assume that the a priori information constraints parameter estimates to the following ranges:
\begin{equation}\label{eq:constraints}
\begin{aligned}
& \ell_{n,1}\leq \hat{\theta}_{n,1} \leq up_{n,1}  \hspace{-1.5cm} &&  \hspace{-1.5cm} \ell_{n,2}\leq \hat{\theta}_{n,2} \leq up_{n,2}\\ 
&& \ell_{n,3}\leq \hat{\theta}_{n,3} \leq up_{n,3}.
\end{aligned}
\end{equation}
Observe that the parameters $\rho_{1},\rho_{2} \in \mathbb{R}^{+}$ have to be tuned. To assess how the choice of these two parameters affects the satisfaction of \eqref{eq:constraints}, consider the number of steps the local estimates violate the constraints over the estimation horizon $T$, $\{{N}_{i}^{b}\}_{i=1}^{3}$. Assuming that \textquotedblleft negligible\textquotedblright \ violations of the constraints are allowed, \eqref{eq:constraints} are supposed to be violated if the estimated parameters fall outside the interval $\mathcal{B}_{n}=\smallmat{\ell_{n}-10^{-4} & u_{n}+10^{-4}}$. Considering the set of constraints
\begin{equation*}
\mathcal{S}_{2}=\{\ell_{n}=\smallmat{0.19 & \theta_{n,2}-0.1 & 0.79}, up_{n}=\smallmat{0.21 & \theta_{n,2}+0.1 & 0.81}\},
\end{equation*}
\figurename{~\ref{Fig:rhostudy_ex1}} shows the average percentage of violations over the $N$ agents obtained fixing $\rho_{2}=0.1$ and choosing 
\begin{equation*}
\rho_{1}=\{10^{-5},10^{-4},10^{-3},10^{-2}.10^{-1},1,10,20\}.
\end{equation*}
Observe that if $\rho_{1}$ dominates over $\rho_{2}$ the number of violations tends to decrease, as in the augmented Lagrangian \eqref{eq:constraints} are weighted more than the consensus constraint. However, if $\rho_{1}/\rho_{2}>100$, $\{\bar{N}_{i}^{b}\}_{i=1}^{3}$ tend to slightly increase. It is thus important to trade-off between the weights attributed to \eqref{eq:constraints} and the consensus constraint.
\begin{figure}[!tb]
	\centering
	\includegraphics[scale=0.7]{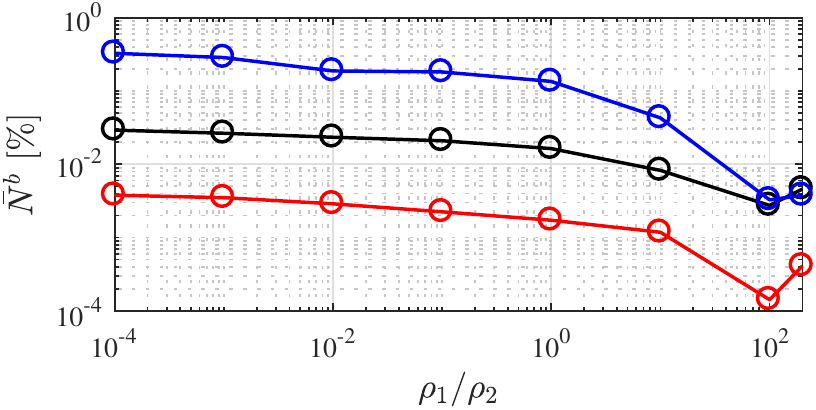}
	\caption{Example 4. $\bar{N}^{b}$ vs $\rho_{1}/\rho_{2}$: black = $\bar{N}_{1}^{b}$, red = $\bar{N}_{2}^{b}$, blue = $\bar{N}_{3}^{b}$.}
	\label{Fig:rhostudy_ex1}
\end{figure}
To evaluate how the stiffness of the constraints affects the choice of the parameters, $\{N_{i}^{b}\}_{i=1}^{3}$ are computed considering three different sets of box constraints
\begin{align*}
 \mathcal{S}_{1}&=\{\ell_{n}=\smallmat{0.195 & \theta_{n,2}-0.05 & 0.795}, up_{n}=\smallmat{0.205 & \theta_{n,2}+0.05 & 0.805}\},\\ \mathcal{S}_{2}&=\{\ell_{n}=\smallmat{0.19 & \theta_{n,2}-0.1 & 0.79}, up_{n}=\smallmat{0.21 & \theta_{n,2}+0.1 & 0.81}\},\\ \mathcal{S}_{3}&=\{\ell_{n}=\smallmat{0.15 & \theta_{n,2}-0.5 & 0.75}, up_{n}=\smallmat{0.25 & \theta_{n,2}+0.5 & 0.85}\}.
\end{align*}
The resulting $\{\bar{N}_{i}^{b}\}_{i=1}^{3}$ are reported in \figurename{~\ref{Fig:rhostudy2_ex1}}.
\begin{figure}[!tb]
	\centerline{
		\begin{tabular}[!tb]{cc}
			\subfigure[$\bar{N}_{1}^{b}$]{\includegraphics[scale=0.7]{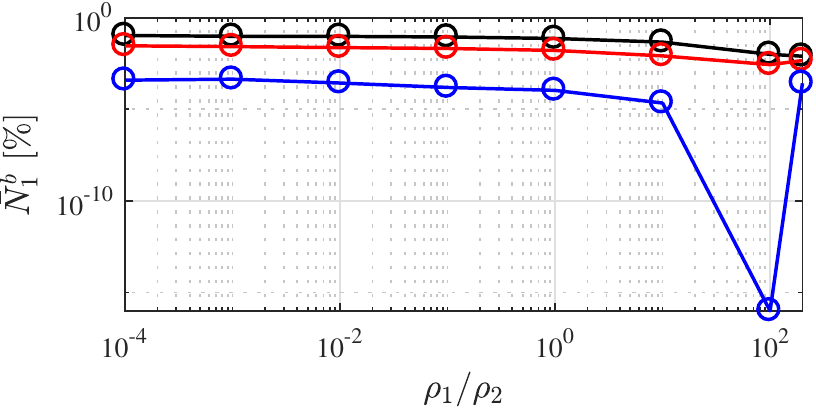}}
			\subfigure[$\bar{N}_{2}^{b}$]{\includegraphics[scale=0.7]{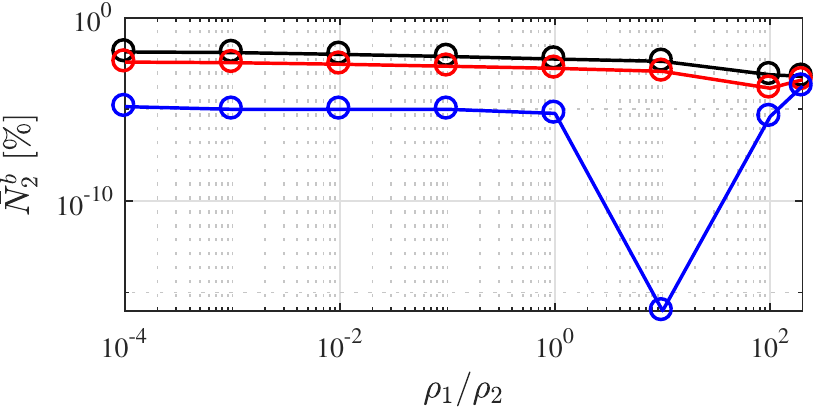}}\\
			\multicolumn{1}{c}{\subfigure[$\bar{N}_{3}^{b}$]{\includegraphics[scale=0.7]{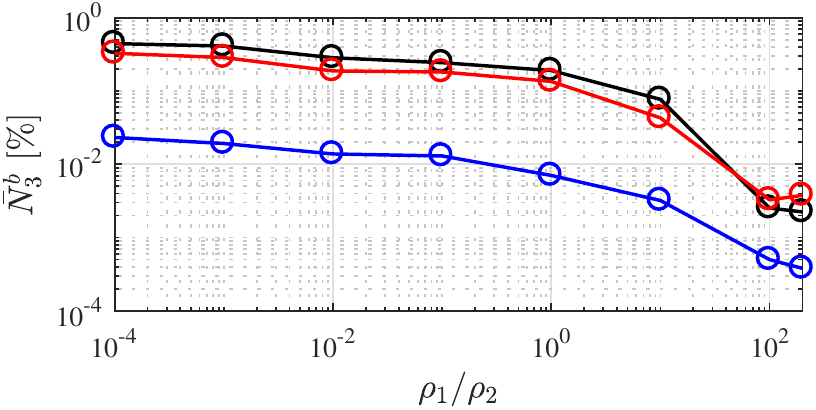}}}\\
		\end{tabular}
	}
	\caption{Example 4. Average percentage of constraint violations $\bar{N}_{i}^{b}$ \%, $i=1,2,3$, vs $\rho_{1}/\rho_{2}$. Black : $\mathcal{S}_{1}$, red : $\mathcal{S}_{2}$, blue : $\mathcal{S}_{3}$.}
	\label{Fig:rhostudy2_ex1}
\end{figure}
Note that also in this case the higher the ratio $\rho_{1}/\rho_{2}$ is, the smaller $\{\bar{N}_{i}^{b}\}_{i=1}^{3}$ are. However, also in this case, the constraint violations tend to increase for $\rho_{1}/\rho_{2}>100$.\\

Focusing on the assessment of ADMM-RLS performances when the set of constraints is $\mathcal{S}_{2}$, \figurename{~\ref{Fig:glob_ex4}} shows the global estimates obtained using the same initial conditions and forgetting factors as in Section~\ref{Sec:3}, with $\rho_{1}=10$ and $\rho_{2}=0.1$.
\begin{figure}[!tb]
	\centerline{
		\begin{tabular}[t]{cc}
			\subfigure[$\theta_{1}^{g}$ vs $\hat{\theta}_{1}^{g}$]{\includegraphics[scale=0.7]{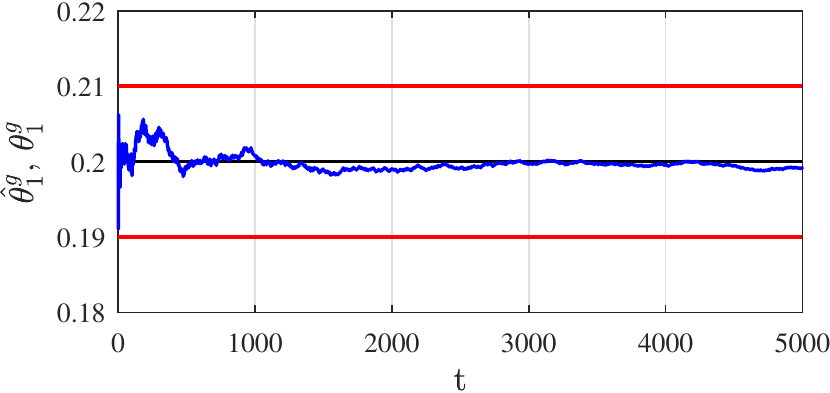}}
			\subfigure[$\theta_{2}^{g}$ vs $\hat{\theta}_{2}^{g}$]{\includegraphics[scale=0.7]{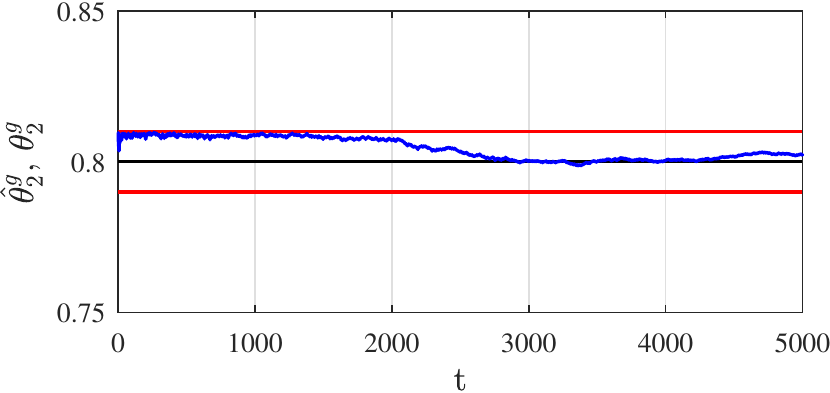}}\\
		\end{tabular}
	}
	\caption{Example 4.Global model parameters: black = true, blue = ADMM-RLS, red = upper and lower bounds.}
	\label{Fig:glob_ex4}
\end{figure}
Note that the global estimates satisfy \eqref{eq:constraints}, showing that the constraints on the global estimate are automatically enforced imposing $\theta_{n} \in \mathcal{C}_{n}$.
As it concerns the RMSEs for $\hat{\theta}^{g}$ \eqref{eq:rmse_glob}, they are equal to:
\begin{align*}
RMSE_{1}^{g}=0.001 \mbox{ and } RMSE_{2}^{g}=0.006,
\end{align*}
and their relatively small values can be related to the introduction of the additional constraints, that allow to limit the resulting estimation error.\\

\figurename{~\ref{Fig:loc_ex4} show the estimate $\hat{\theta}_{n}$ for $n=11$, with $SNR_{11}=10.6$~dB. Note that the estimated parameters tend to satisfy the constraints.
	\begin{figure}[!tb]
		\centerline{
			\begin{tabular}[t]{cc}
				\subfigure[$\theta_{11,1}$ vs $\hat{\theta}_{11,1}$]{\includegraphics[scale=0.7]{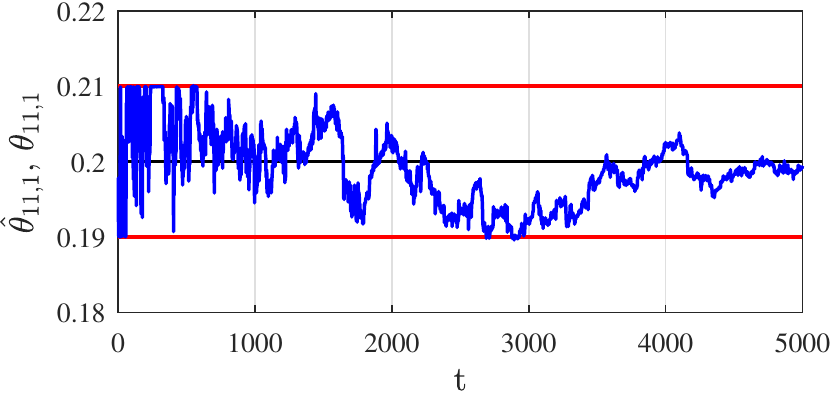}}
				\subfigure[$\theta_{11,2}$ vs $\hat{\theta}_{11,2}$]{\includegraphics[scale=0.7]{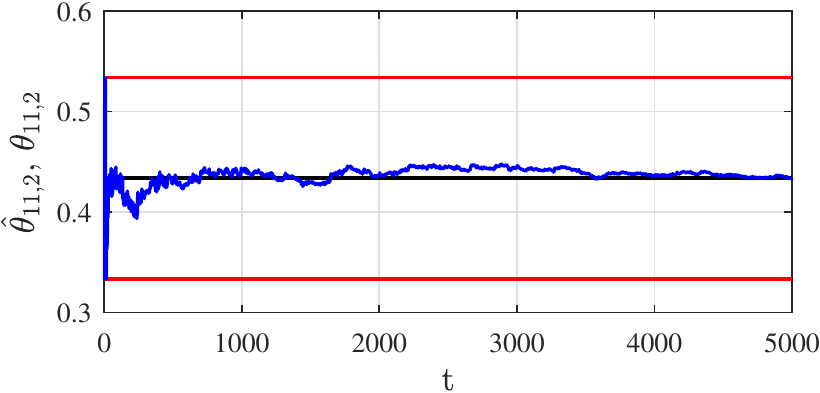}}\\
				\multicolumn{1}{c}{\subfigure[$\theta_{11,3}$ vs $\hat{\theta}_{11,3}$]{\includegraphics[scale=0.7]{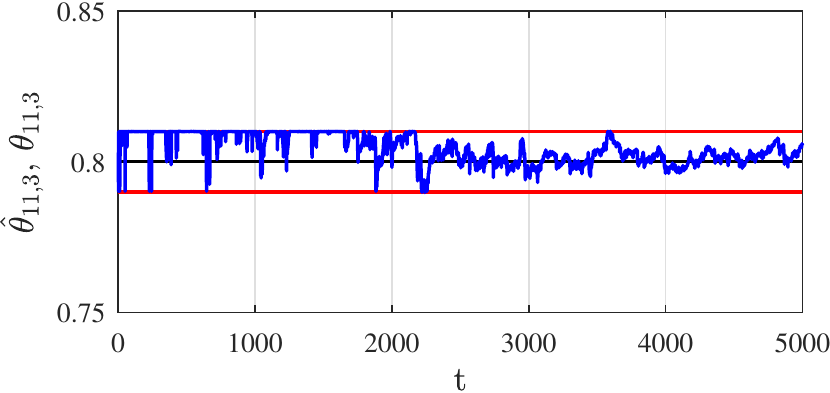}}}\\
			\end{tabular}
		}
		\caption{Local parameter $\theta_{n}$, $n=11$. Black : true, blue : ADMM-RLS, red : upper and lower bounds}
		\label{Fig:loc_ex4}
	\end{figure}
In \figurename{~\ref{Fig:locvsrls_ex4}} $\hat{\theta}_{n}$ and $\hat{\theta}_{n}^{RLS}$, with $n=11$, are compared. As it can be noticed, while $\hat{\theta}_{11}$ satisfied the imposed constraints on its values, the effect of using $\hat{\theta}_{11}$ to update $\hat{\theta}_{11}^{RLS}$ (see~\eqref{eq:rls_parC}) is not strong enough to enfoce also the estimates computed locally to satisfy the contraints.
	\begin{figure}[!tb]
	\centerline{
		\begin{tabular}[t]{cc}
			\subfigure[$\theta_{11,1}$ vs $\hat{\theta}_{11,1}$ and $\hat{\theta}_{11,1}^{RLS}$]{\includegraphics[scale=0.7]{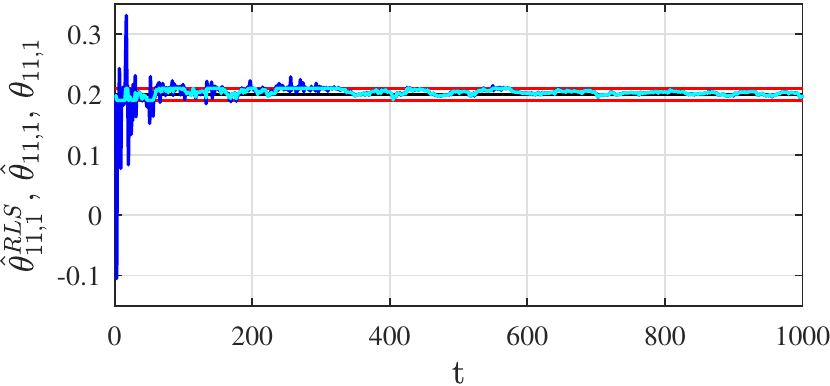}}
			\subfigure[$\theta_{11,2}$ vs $\hat{\theta}_{11,2}$ and $\hat{\theta}_{11,2}^{RLS}$]{\includegraphics[scale=0.7]{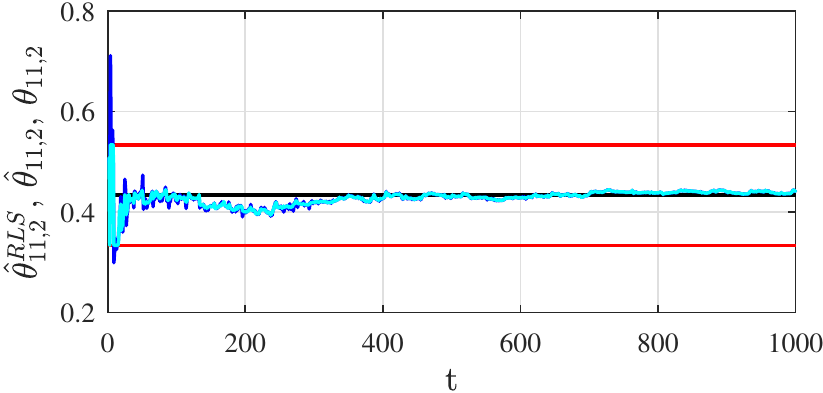}}\\
			\multicolumn{1}{c}{\subfigure[$\theta_{11,3}$ vs $\hat{\theta}_{11,3}$ and $\hat{\theta}_{11,3}^{RLS}$]{\includegraphics[scale=0.7]{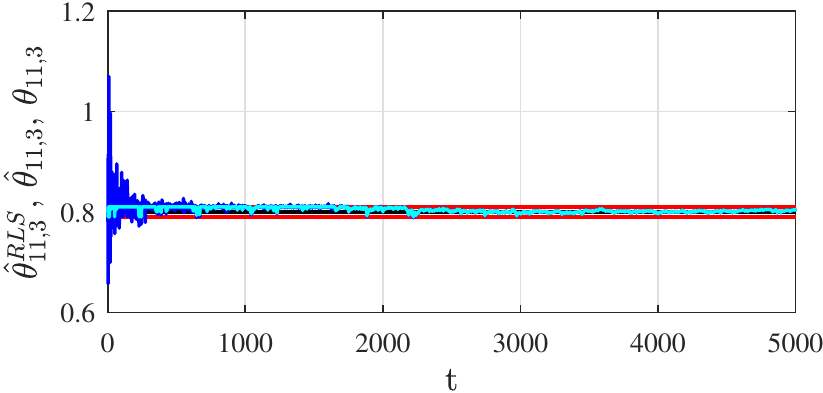}}}\\
		\end{tabular}
	}
	\caption{Local parameter $\theta_{n}$, $n=11$, for $t \in [1 \ 1000]$. Black : true, blue : $\hat{\theta}_{11}^{RLS}$, cyan : $\hat{\theta}_{11}$, red : upper and lower bounds}
	\label{Fig:locvsrls_ex4}
\end{figure}
	To further assess the performance of the proposed approach, the RMSE for the local estimates 
	\begin{equation}\label{eq:rmse_loc}
	\mathrm{RMSE}_{n,i}=\sqrt{\frac{\sum_{t=1}^T\left(\theta_{n,i}-\hat{\theta}_{n,i}(t)\right)^2}{T}}.
	\end{equation}
	is also considered. $RMSE_{n,2}$ obtained for each of the $N$ systems is reported in \figurename{~\ref{Fig:rmse_ex4}} and, as it can be noticed, $RMSE_{n,2}$ is relatively small. As for the global parameters' estimates, this result can be related to the introduction of the additional constraints.
	\begin{figure}[!tb]
		\centering
		\includegraphics[scale=0.7]{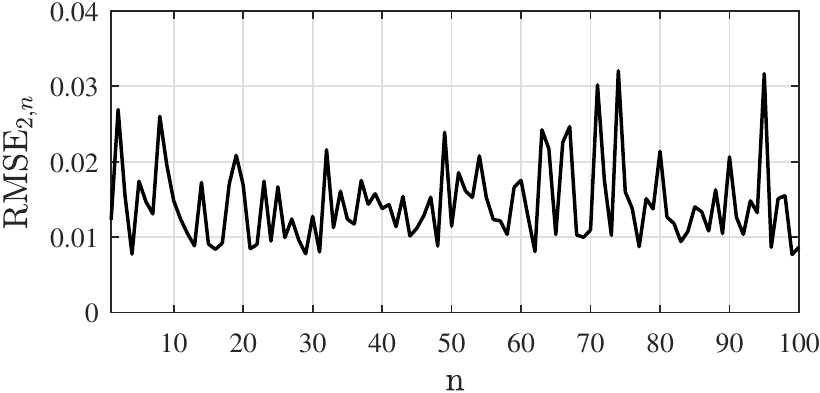}
		\caption{$RMSE_{2}$ for each agent $n$, $n=1,\ldots,N$.}
		\label{Fig:rmse_ex4}
	\end{figure}
\section{Concluding Remarks and Future Work}\label{Sec:Conclusions}
In this report a method for collaborative least-squares parameter estimation is presented based on output measurements from multiple systems which can perform local computations and are also connected to a centralized resource in the \textquotedblleft cloud\textquotedblright. The approach includes two stages: ($i$) a \emph{local} step, where estimates of the unkown parameters are obtained using the locally available data, and ($ii$) a \emph{global} stage, performed on the cloud, where the local estimates are fused.\\
Future research will address extentions of the method to the nonlinear and multi-class consensus cases. Moreover, an alternative solution of the problem will be studied so to replace the transmission policy required now, i.e. N2C2N, with a Node-to-Cloud (N2C) communication scheme. This change should allow to alleviate problems associated with the communication latency between the cloud and the nodes. Moreover, it should enable to obtain local estimators that run independently from the data transmitted by the cloud, and not requiring synchronous processing by the nodes and \textquotedblleft cloud\textquotedblright. Other, solutions to further reduce the trasmission complexity and to obtain an asynchronous scheme with the same characteristics as the one presented in this report will be investigated.
\begin{appendices}
	\section{Centralized RLS}\label{Appendix:A}
	Consider problem \eqref{eq:gen_consensus}, with the cost functions given by
	\begin{equation*}
	f_{n}(\theta_{n})=\frac{1}{2}\sum_{t=1}^{T}\|y_{n}(t)-(X_{n}(t))'\theta_{n}\|_{2}^{2}.
	\end{equation*}
	The addressed problem can be solved in a fully centralized fashion, if at each step $t$ all the agents transmit the collected data pairs $\{y_{n}(t),X_{n}(t)\}$, $n=1,\ldots,N$, to the \textquotedblleft cloud\textquotedblright. This allows the creation of the lumped measurement vector and regressor, given by
	\begin{equation}
	\begin{aligned}
	\check{y}(t)&=\begin{bmatrix}y_{1}(t)' & \ldots & y_{N}(t)'\end{bmatrix}' \in \mathbb{R}^{N \cdot n_{y} \times 1 }, \\
	\check{X}(t)&=\begin{bmatrix}X_{1}(t)' & \ldots & X_{N}(t)'\end{bmatrix}' \in \mathbb{R}^{ n_{\theta}\times n_{y} \cdot N}.
	\end{aligned}
	\end{equation} 
	Through the introduction of the lumped vectors, \eqref{eq:gen_consensus} with $f_{n}$ as in \eqref{eq:least_sqCost} is equivalent to
	\begin{equation}\label{eq:c_RLS}
	\min_{\theta^{g}} \frac{1}{2}\sum_{t=1}^{T} \left\|\check{y}(t)-(\check{X}(t))'\theta^{g}\right\|_{2}^{2}.
	\end{equation}
	The estimate for the unknown parameters $\hat{\theta}^{g}$ can thus be retrieved applying standard RLS (see \cite{ljung1999system}), i.e. performing at each step $t$ the following iterations
	\begin{align}
	\mathcal{K}(t)&=\phi(t-1)\check{X}(t)\left(I_{\hspace{-0.1cm}\footnotesize{\mbox{\textcalligra{D}}}\hspace{0.1cm}}+(\check{X}(t))'\phi(t-1)\check{X}(t)\right)^{-1},\\	\phi(t)&=\left(I_{n_{\theta}}-\mathcal{K}(t)(\check{X}(t))'\right)\phi(t-1),\\
	\hat{\theta}^{g}(t)&=\hat{\theta}^{g}(t-1)+\mathcal{K}(t)\left(\check{y}(t)-(\check{X}(t))'\hat{\theta}^{g}(t-1)\right),
	\end{align}
	with \textcalligra{D} $=N \cdot n_{y} \times 1$.
\end{appendices}


\bibliographystyle{plain}
\bibliography{CloudAided_linear}

\begin{thebibliography}{10}

\bibitem{Boem2012}
F.~Boem, Y.~Xu, C.~Fischione, and T.~Parisini.
\newblock A distributed estimation method for sensor networks based on pareto
  optimization.
\newblock In {\em 2012 IEEE 51st IEEE Conference on Decision and Control
  (CDC)}, pages 775--781, Dec 2012.

\bibitem{ADMMBoyd}
S.~Boyd, N.~Parikh, E.~Chu, B.~Peleato, and J.~Eckstein.
\newblock Distributed optimization and statistical learning via the alternating
  direction method of multipliers.
\newblock {\em Found. Trends Mach. Learn.}, 3(1):1--122, January 2011.

\bibitem{Cattivelli2008}
F.~S. Cattivelli, C.~G. Lopes, and A.~H. Sayed.
\newblock Diffusion recursive least-squares for distributed estimation over
  adaptive networks.
\newblock {\em IEEE Transactions on Signal Processing}, 56(5):1865--1877, May
  2008.

\bibitem{Forero2010}
Pedro~A. Forero, Alfonso Cano, and Georgios~B. Giannakis.
\newblock Consensus-based distributed support vector machines.
\newblock {\em The Journal of Machine Learning Research}, 11:1663--1707, Aug
  2010.

\bibitem{Garin2010}
F.~Garin and L.~Schenato.
\newblock {\em A Survey on Distributed Estimation and Control Applications
  Using Linear Consensus Algorithms}, pages 75--107.
\newblock Springer London, London, 2010.

\bibitem{howell2010brake}
M.N. Howell, J.P. Whaite, P.~Amatyakul, Y.K. Chin, M.A. Salman, C.H. Yen, and
  M.T. Riefe.
\newblock Brake pad prognosis system, Apr 2010.
\newblock US Patent 7,694,555.

\bibitem{Li2016}
Z.~Li, I.~Kolmanovsky, E.~Atkins, J.~Lu, D.~P. Filev, and J.~Michelini.
\newblock Road risk modeling and cloud-aided safety-based route planning.
\newblock {\em IEEE Transactions on Cybernetics}, 46(11):2473--2483, Nov 2016.

\bibitem{Li2017}
Z.~Li, I.~Kolmanovsky, E.~M. Atkins, J.~Lu, D.~P. Filev, and Y.~Bai.
\newblock Road disturbance estimation and cloud-aided comfort-based route
  planning.
\newblock {\em IEEE Transactions on Cybernetics}, PP(99):1--13, 2017.

\bibitem{ljung1999system}
L.~Ljung.
\newblock {\em {System identification: theory for the user}}.
\newblock Prentice-Hall Englewood Cliffs, NJ, 1999.

\bibitem{Lopes2007}
C.~G. Lopes and A.~H. Sayed.
\newblock Incremental adaptive strategies over distributed networks.
\newblock {\em IEEE Transactions on Signal Processing}, 55(8):4064--4077, Aug
  2007.

\bibitem{Mateos2009}
G.~Mateos, I.~D. Schizas, and G.~B. Giannakis.
\newblock Distributed recursive least-squares for consensus-based in-network
  adaptive estimation.
\newblock {\em IEEE Transactions on Signal Processing}, 57(11):4583--4588, Nov
  2009.

\bibitem{Mell2011}
Peter~M. Mell and Timothy Grance.
\newblock Sp 800-145. the nist definition of cloud computing.
\newblock Technical report, Gaithersburg, MD, United States, 2011.

\bibitem{Olfati-Saber2007}
R.~Olfati-Saber.
\newblock Distributed kalman filtering for sensor networks.
\newblock In {\em 2007 46th IEEE Conference on Decision and Control}, pages
  5492--5498, Dec 2007.

\bibitem{Ozatay2014}
E.~Ozatay, S.~Onori, J.~Wollaeger, U.~Ozguner, G.~Rizzoni, D.~Filev,
  J.~Michelini, and S.~Di Cairano.
\newblock Cloud-based velocity profile optimization for everyday driving: A
  dynamic-programming-based solution.
\newblock {\em IEEE Transactions on Intelligent Transportation Systems},
  15(6):2491--2505, Dec 2014.

\bibitem{Taheri2016}
E.~Taheri, O.~Gusikhin, and I.~Kolmanovsky.
\newblock Failure prognostics for in-tank fuel pumps of the returnless fuel
  systems.
\newblock In {\em Dynamic Systems and Control Conference}, Oct 2016.

\end{thebibliography}

\end{document}